\DocumentMetadata{}
\documentclass[screen, nonacm]{acmart}
\AtBeginDocument{%
  \providecommand\BibTeX{{%
    \normalfont B\kern-0.5em{\scshape i\kern-0.25em b}\kern-0.8em\TeX}}}

\usepackage{enumitem} 
\usepackage{algorithm} 
\usepackage{bm} 
\usepackage{setspace} 
\usepackage{stfloats} 
\usepackage{url} 
\usepackage{algpseudocode}
\usepackage[normalem]{ulem}
\usepackage{listings}
\usepackage{xcolor}
\definecolor{codegreen}{rgb}{0,0.6,0}
\definecolor{codegray}{rgb}{0.5,0.5,0.5}
\definecolor{codepurple}{rgb}{0.58,0,0.82}
\definecolor{backcolour}{rgb}{0.95,0.95,0.92}

\lstdefinestyle{mystyle}{
  backgroundcolor=\color{backcolour}, commentstyle=\color{codegreen},
  keywordstyle=\color{codegreen},
  numberstyle=\tiny\color{codegreen},
  stringstyle=\color{codegreen},
  basicstyle=\ttfamily\color{codegreen}\footnotesize,
  breakatwhitespace=false,         
  breaklines=true,                 
  captionpos=b,                    
  keepspaces=true,                 
  numbers=left,                    
  numbersep=5pt,                  
  showspaces=false,                
  showstringspaces=false,
  showtabs=false,                  
  tabsize=2
}

\begin{document}
\title{Enabling Generative Design Tools with LLM Agents for Mechanical Computation Devices: A Case Study}

\author{Qiuyu Lu}
\authornote{Contributed equally}
\orcid{0000-0002-8499-3091}
\affiliation{%
  \institution{University of California, Berkeley}
  \streetaddress{}
  \city{California}
  \state{CA}
  \country{USA}
  \postcode{94720}
}
\email{qiuyulu@berkeley.edu}

\author{Jiawei Fang}
\authornotemark[1]
\affiliation{%
  \institution{University of California, Berkeley}
  \city{California}
  \state{CA}
  \country{USA}}
\email{jiaweif@berkeley.edu}

\author{Zhihao Yao}
\authornotemark[1]
\affiliation{%
  \institution{Tsinghua University}
  \city{Beijing}
  \country{China}}
\email{yaozh_h@outlook.com}

\author{Yue Yang}
\affiliation{%
  \institution{University of California, Berkeley}
  \city{California}
  \state{CA}
  \country{USA}}
 \email{yue.yang@berkeley.edu}

\author{Shiqing Lyu}
\affiliation{%
  \institution{Tsinghua University}
  \city{Beijing}
  \country{China}}
\email{lvsq22@mails.tsinghua.edu.cn}

\author{Haipeng Mi}
\affiliation{%
  \institution{Tsinghua University}
  \city{Beijing}
  \country{China}}
\email{mhp@tsinghua.edu.cn}

\author{Lining Yao}
\affiliation{%
  \institution{University of California, Berkeley}
  \streetaddress{}
  \city{California}
  \state{CA}
  \country{USA}}
\email{liningy@berkeley.edu}

\renewcommand{\shortauthors}{Lu, et al.}

\begin{abstract}

In the field of Human-Computer Interaction (HCI), interactive devices with embedded mechanical computation are gaining increasing attention. The rise of these cutting-edge devices has highlighted the need for specialized design tools that democratize the prototyping process. While current tools streamline the process through parametric design and simulation, they often come with a steep learning curve and may not fully support creative ideation. In this study, we use fluidic computation interfaces as a case study to explore how design tools for such devices can be augmented by Large Language Model agents (LLMs). Integrated with LLMs, the Generative Design Tool (GDT) better understand the capabilities and limitations of new technologies, propose diverse, practical application, and suggest designs that are technically and contextually appropriate. Additionally, it generates design parameters for visualizing results and producing fabrication-ready support files. This paper details the GDT’s framework, implementation, and performance, while addressing its potential and challenges.

\end{abstract}




\maketitle

\section{Introduction}

Interactive devices with embedded mechanical computation are gaining increasing attention in HCI~\cite{savage_airlogic_2022, ion_digital_2017, lu_fluidic_2023, bonbon}. As novel mechanical computation techniques emerge, prototyping these devices often requires highly specialized expertise. To address this challenge, specialized design tools have been developed to lower the barrier through parametric design and simulation~\cite{savage_airlogic_2022,ion_digital_2017}. However, designing such devices remains complex. Designers must have a comprehensive understanding of the new technology’s capabilities and limitations, as well as the implications of all design parameters. Additionally, generating appropriate application ideas requires extensive exploration to meet various requirements, from functional goals to feasibility constraints. Even experts in mechanical device design, including developers and researchers of the tools themselves, must invest considerable effort to propose diverse use cases that demonstrate the technology’s application value.


In universal design software, like computer-aided design (CAD) software, artificial intelligence (AI) integration simplifies complex tasks and optimizes structures and materials autonomously \cite{miller2019explanation, dataCAD}. However, incorporating AI features often requires extensive algorithm development or substantial data training \cite{willis2022joinable, zhao2020robogrammar}, which is impractical for specialized design tools without significant effort. Moreover, users still need domain knowledge to define the design problem, set goals, and adjust parameters before benefiting from AI, leaving the original challenge unresolved.


Large Language Models (LLMs) like GPT excel in creative tasks due to their vast knowledge, nuanced language generation, and contextual awareness \cite{PromptSapper,UIllm,CAM,t2image}. For example, 3DALL-E \cite{liu20233dall} introduced a Fusion 360 plugin that uses GPT to generate text and image prompts for modeling, leveraging its extensive design knowledge. Recently, LLM agents have garnered attention for their ability to learn new information and autonomously handle complex tasks \cite{auto2023, baby2023, yao2022react, park2023generative, li2024camel, hong2023metagpt}. This leads to an intriguing question: can a synergy between LLMs and specialized design tools revolutionize the prototyping of novel devices by harnessing LLMs’ flexibility, adaptability, and creative problem-solving?


To investigate how LLMs can enhance the design workflows of mechanical computation devices, we developed a generative design tool (GDT) augmented with LLM agents, focusing on the fluidic computation interface (FCI) proposed by Lu et al. \cite{lu_fluidic_2023}. FCI was selected as a case study for several reasons: 1) it qualifies as a mechanical computation device; 2) it requires specialized domain knowledge not typically available in general-purpose LLMs; 3) it integrates diverse inputs, outputs, and computational functions, reflecting the complexity of many interactive devices; and 4) it thoroughly explores its design space, providing a strong foundation for refining LLM agents with relevant knowledge and logic.



The tool presented here assists designers with FCI-related inquiries and offers inspiration for various application scenarios. When a designer inputs a vague goal (e.g., creating a “smart Yoga pad”), the tool provides detailed design suggestions. It guides the user through refining the design, from selecting functional modules to setting parameters, by prompting with recommendations and responding to user queries. Additionally, the tool verifies the final design with interaction simulations and supports the physical device’s construction by generating fabrication files. We evaluate its performance in proposing and realizing DGs, reflecting on the benefits and challenges of integrating LLM agents into specialized design tools. We hope this work inspires researchers and broadens the user base for such tools in novel device design.


The contributions of this work are as follows:
\begin{itemize}
\item Developing an approach for LLM-enhanced design tools for mechanical computation devices (specifically FCI), using a multi-agent framework and straightforward adjustments to equip LLM agents with FCI-specific design knowledge.
\item Creating the GDT for FCI, integrating LLM agents to apply fluid computation in designing novel interactive devices.
\item Evaluating the GDT’s performance and reflecting on the benefits and challenges of incorporating LLM agents into specialized design tools for building novel devices.
\end{itemize}


\section{Related Work}

\subsection{AI Enhanced Design Tools for Building Devices}

Specialized design tools are essential for prototyping novel devices \cite{metamaterial, DMM, Sustainflatable, ou_aeromorph_2016}. AI integration has transformed the design process, especially in physical prototyping, by reducing trial-and-error costs, boosting efficiency, and expanding design possibilities \cite{gmeiner2023exploring}. For example, Zanzibar facilitates rapid game development by exploring the design space of physical and digital games \cite{Zanzibar}, while 3D and 4D printing offer intuitive simulations and support interactive design processes \cite{Thermorph,RoMA,SimuLearn,4Doodle,XBridges,ShapeAware}. These technologies are widely applied in areas such as shape-changing materials \cite{EpoMemory} and circuit design \cite{Toastboard,PrintScreen}. However, despite the efficiency gains from AI-integrated tools, challenges remain, including limited inspiration, reliance on users to set numerous parameters, and a lack of contextual awareness and proactive suggestions \cite{gmeiner2023exploring, li2024study}.

Meanwhile, LLMs have shown promise in enhancing creative tools across various domains \cite{PromptSapper,UIllm,CAM,t2image,liu20233dall}, allowing designers to start with simple requirements and complete projects with LLM assistance. However, in the field of designing interactive hardware systems—which involve input, computation, and output functionalities, and interact contextually with users—LLM integration remains limited. This is likely because LLMs are trained on large-scale textual data and lack the specialized knowledge required for interactive hardware design. These systems demand complex combinations of components, structured information, and high precision to ensure functional physical implementation, which can be difficult to describe using simple text.

Nevertheless, we envision there is a chance to harness LLM's capabilities to understand and innovate new interactive device technologies. By augmenting conventional specialized design tools with LLM, we can introduce a fresh approach to device prototyping. Leveraging LLM's strengths in flexibility, adaptability, and creative problem-solving can lower usage barriers and inspire more innovative designs.


\subsection{Large Language Models and Agents}

Large Language Models (LLMs), trained on vast datasets to handle billions of parameters (e.g., GPT-4 \cite{achiam2023gpt, andreas2022language}), have become pivotal with the rise of Transformers \cite{vaswani2017attention}, enabling them to generate human-like text. Their powerful capabilities have found broad application in Human-Computer Interaction (HCI), enhancing user interaction \cite{liu2023wants, dang2023choice, jiang2023graphologue, zamfirescu2023johnny, suh2023sensecape, wang2023reprompt} and supporting design processes in CAD and animation \cite{liu20233dall, tseng2024keyframer}. Additionally, LLMs have been studied for ethical concerns, such as misinformation risks \cite{zhou2023synthetic}, and have been applied in art, public health, journalism, and education \cite{valencia2023less, huh2023genassist, jo2023understanding, petridis2023anglekindling, lu2023readingquizmaker}. However, we still have yet to see the LLM-based generative design tools being fully leveraged to design highly specialized interactive devices.

An agent refers to an entity capable of autonomously performing tasks, making decisions, and interacting with its environment or other agents to achieve specific goals \cite{jennings1998roadmap,vere1990basic}. In the LLM Agent framework, the LLM acts as the agent’s brain, providing core functions like reasoning, planning, and decision-making \cite{wang2023survey}. The agent also integrates long-term memory through external knowledge libraries for quick retrieval, short-term memory via in-context learning, and tool use through APIs. These features allow LLM agents to tackle complex tasks efficiently across diverse applications. Research in this area is divided between single-agent and multi-agent approaches. Early work focused on single agents using tools to address complex tasks \cite{baby2023,yao2022react,auto2023}, but this often led to hallucinations and task failures \cite{chen2023agentverse}. The multi-agent approach, which decomposes tasks into subtasks handled by specialized agents, improves accuracy and completeness by leveraging each agent’s expertise \cite{chen2023autoagents,chan2023chateval,park2023generative}. This collaborative method significantly enhances the ability to solve complex tasks compared to single-agent systems \cite{li2024camel,li2023metaagents}. In our work, we adopted a multi-agent strategy to lay the foundation for our tool’s design (Fig. \ref{fig:architecture}).

\subsection{Mechanical Computation Devices}
Mechanical computation has garnered increasing attention in recent years, marking its success across various science and engineering fields such as molecular computation \cite{1.4}, robotics \cite{1.5,oct}, morphological computation \cite{1.7}, and fluidic logic circuits \cite{Next-generation,Logic-digital-fluidic,1st-pneumatic-logic-circuits, preston_soft_2019, rajappan_logic-enabled_2022, preston_digital_2019, shveda_wearable_2022}. Within the HCI device fabrication and design community, there's a notable trend toward leveraging unconventional computation to enhance computing capabilities. A handful of HCI studies have explored the integration of mechanical computation within interfaces. Ion et al. introduced digital mechanical metamaterials for constructing logic structures \cite{metamaterial}. Venous Materials \cite{venous} showcased fluidic mechanisms that react to mechanical inputs from users to produce outputs. Logic Bonbon \cite{bonbon} created basic logic gates for crafting desserts with varying flavors based on user inputs. AirLogic \cite{savage_airlogic_2022} advanced these fluidic logic concepts by incorporating them into 3D printing processes. 
The Fluidic Computation Kit \cite{lu_fluidic_2023} delves deeper, offering a systematic framework for developing advanced mechanical computation devices, incorporating diverse force inputs, computation operators, and output modalities. Its broad scope makes it especially compelling, as it covers multiple aspects of interactive device construction beyond just actuation or sensing. In addition, Due to its complexity and innovation, the technology poses unique challenges for LLM agents. Thus, we selected the FCI as our case study to explore the potential and limitations of LLM agents in enhancing novel device design tools.

\section{Background Knowledge on FCI}

\begin{figure}[]
    \centering
    \includegraphics[width=0.5\linewidth]{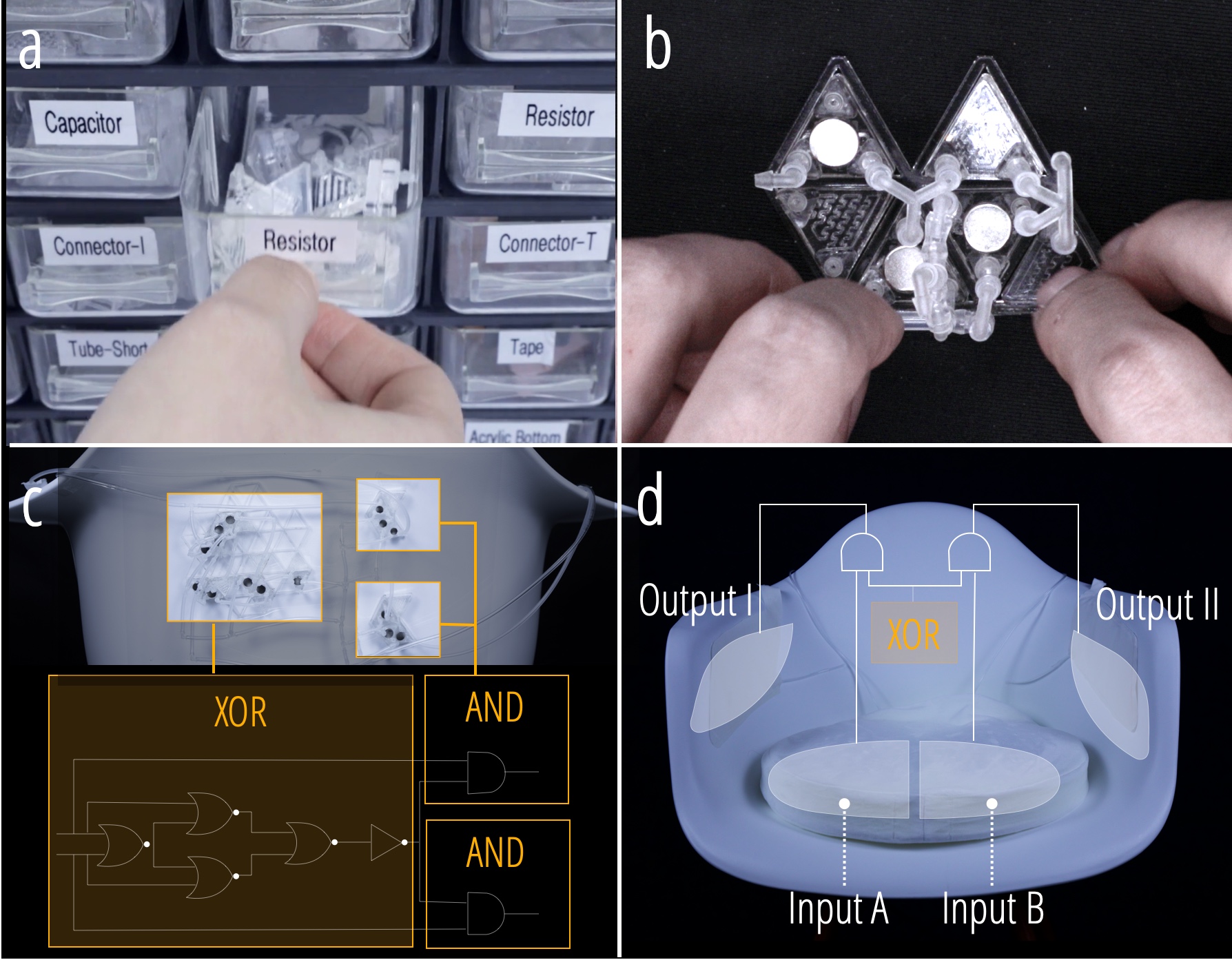}
    \caption{
    Building an interactive sitting posture correction chair using the Fluidic Computation Kit involves: a) Selecting the basic components; b) Assembling operators by wiring the components; c) Constructing the circuit with operators based on the logic; and d) Preparing and integrating input/output airbags with the circuit into the chair. (Permission granted from the authors)}
    \label{fig:background}
\end{figure}

In our context, FCI refers to devices built using the Fluidic Computation Kit (Fig. \ref{fig:background}) \cite{lu_fluidic_2023}, which utilizes fluidic circuits with pneumatic components like valves to perform logic computations via pneumatic signals. These signals, represented by positive pressure (1) and atmospheric pressure (0), drive logic operations and generate outputs. The design space of FCI encompasses three key elements: force input, mechanical computation, and tangible output.

\begin{figure*}[]
  \includegraphics[width=1\linewidth]{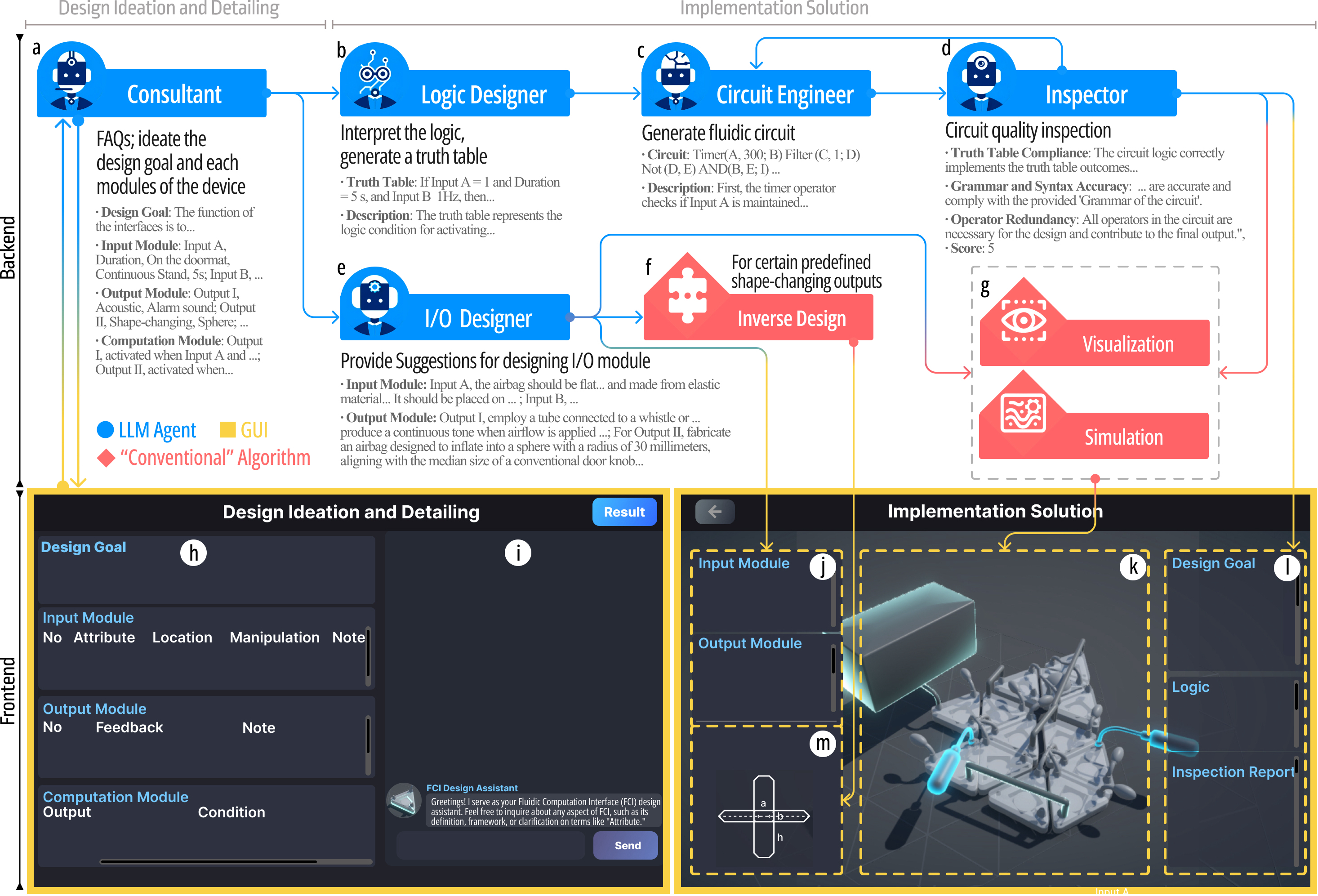}
  \caption{The GDT’s overview. Sections a-g constitute the backend, which includes LLM agents and conventional algorithms. Sections h-m represent the frontend, utilized by users interacting with the GDT. The \textit{Consultant} agent (a) and the first tab of the GUI (h,j) are dedicated to assisting with the design ideation and detailing phase. The remaining components are focused on the solution implementation phase. Some text within the GUI has been enlarged to enhance readability. The heat-sealing pattern for the output module (m) will be provided when the type of output is shape-changing, and the shape is associated with an inverse design algorithm.}
  \Description{}
  \label{fig:architecture}
\end{figure*}




- Input: This part consists of airbags that detect changes in internal air pressure caused by external forces. These changes shift the input signal between atmospheric (0) and positive (1), enabling detection of force presence, duration, frequency, and sudden variations.

- Computation: The fluidic circuit processes air pressure inputs through predetermined logic, generating output signals. It uses basic computational components (in triangle form) and operators created by combining these components.

- Output: The output provides feedback in various forms, such as shape-change, haptic, olfactory, and acoustic feedback, all driven by airflow.

\section{Designing with the GDT} 

\begin{figure*}[]
    \centering
    \includegraphics[width=1\linewidth]{Figure/WalkthroughV2.jpg}
    \caption{The user flow for designing with Fluid Computation GDT includes: (1) Greeting the \textit{Consultant} and freely asking for explanations to learn about FCI. (2) Setting the design goal with the help of ideation. (3) Defining the input, output, and computation modules with the \textit{Consultant}'s guidance and recommendations. (4) Confirming the design definition and previewing the generated fluid computation system; clicking the input module to see the animated demonstration of control logic.}
    \label{fig:Walkthrough}
\end{figure*}
\subsection{Overview}


The GDT’s architecture, shown in Fig. \ref{fig:architecture}, includes typical elements of specialized design tools, such as parametric design, rendering previews, and simulations. However, unlike conventional tools that completely rely on users to create design goals (DGs) and manually adjust parameters or use drag-and-drop modules, this GDT, enhanced with LLM agents, can assist or even automate these tasks. The following LLM agents are integrated into this GDT:


\textbf{- \textit{Consultant}} (Fig. \ref{fig:architecture}.a): This agent possesses comprehensive knowledge of FCI’s capabilities and limitations. It answers FCI-related questions, helps define DGs, and assists in deciding design details. Users interact with it via a GUI dialogue box (Fig. \ref{fig:architecture}.i), and it provides real-time updates on the GUI (Fig. \ref{fig:architecture}.h). The DG will also be displayed in the solution window (Fig. \ref{fig:architecture}.l).


\textbf{- \textit{Logic Designer, Circuit Engineer, Inspector}} (Fig. \ref{fig:architecture}.b-d): These agents manage the design of the computation module. The \textit{Logic Designer} analyzes details from the \textit{Consultant} and produces computational logic in a standard format. The \textit{Circuit Engineer}, familiar with FCI operators, selects the appropriate ones to complete the circuit design. The \textit{Inspector} checks the circuit for correctness and conciseness, determining if it needs revisions. Approved designs are visualized and simulated on the GUI (Fig. \ref{fig:architecture}.g, k), with users referencing the visualization to build the FCI circuit. Input and output modules are qualitatively visualized. The logic and inspection report will be displayed at Fig. \ref{fig:architecture}.l.


\textbf{- \textit{I/O Designer}} (Fig. \ref{fig:architecture}.e): This agent designs the I/O module, including components like airbags for detecting inputs or providing output feedback. Based on details from the \textit{Consultant} and its knowledge of the FCI, the \textit{I/O Designer} offers primarily qualitative design suggestions. For predefined output shapes, such as airbags, it will try to provide quantitative geometry (e.g., sphere diameter). Qualitative suggestions are displayed directly in the GUI (Fig. \ref{fig:architecture}.m), while quantitative recommendations are processed via inverse design algorithms to generate heat-sealing patterns and dimensions for user review (Fig. \ref{fig:architecture}.m).


\subsection{Walkthrough}
In this subsection, we provide a design walkthrough demonstrating the use of the GDT to create a smart yoga pad (Fig. \ref{fig:Walkthrough}).

\textbf{Set Design Goal} (Fig. \ref{fig:Walkthrough}.2). After the introduction, the \textit{Consultant} helps Emily define her DG, reassuring her that detailed ideas are unnecessary since the bot can assist in developing them. Emily expresses interest in creating a smart yoga pad. The \textit{Consultant} proposes ideas like Posture Detection, Sequence Guide, and Breathing Aid. Inspired by these suggestions, Emily decides to proceed with detection and physical assistance for challenging poses, explaining: “If a user can't reach the hand area during a pose, the mat will inflate to support, like a yoga block.” The \textit{Consultant} also suggests detecting specific postures that are difficult for beginners, such as the Triangle Pose. With the goal confirmed, Emily moves to the next step.


\textbf{Define Input Module} (Fig. \ref{fig:Walkthrough}.3). With the DG set, the \textit{Consultant} assists in defining the Input Module for the yoga pad. The Input Modules are airbags that detect external forces, triggering pressure changes that shift the input signal from atmospheric (0) to positive (1). Each input is defined by three main properties along with an optional comment note:


\begin{itemize}
\item \textbf{Attribute}: Defines the signal state as Binary (0 or 1), Duration (time in each state), Frequency (transition rate), and Edge (moment of transition).
\item \textbf{Location}: Specifies where input airbags are placed, e.g., under feet, on vehicle sides, or within a seatbelt.
\item \textbf{Manipulation}: Describes interactions with the airbags, such as squeezing, stepping, pressing, or twisting.
\item \textbf{Note} (Optional): Adds extra details, like specific frequency or duration values.
\end{itemize}


In this walkthrough, the \textit{Consultant} suggests two inputs: Input A to detect hand presence at a specific area on the pad, corresponding to the expected posture, and Input B to detect foot presence, confirming the user is on the pad. Both inputs are initially set to binary. Emily requests duration detection for foot Input to avoid false activation, and the \textbackslash{}textit\{Consultant\} adjusts the setting, asking for Emily’s confirmation. 


\textbf{Define Output Module} (Fig. \ref{fig:Walkthrough}.3). After confirming the input module, the \textit{Consultant} moves to defining the output Module, presenting options like \textbf{shape-changing, haptic, olfactory, and acoustic feedback}, aligned with the capabilities of FCI. Given the goal of providing physical assistance when the user’s hand can not reach the mat, the \textit{Consultant} suggests setting Output I to Shape-changing Feedback, which inflates to form a supportive air block when hand support is needed. Emily is satisfied with the result and confirms it. Additionally, the \textit{Consultant} recommends a box shape output with dimensions of 23 × 15 × 7.5 cm, similar to a yoga brick.


\textbf{Define Computation Module} (Fig. \ref{fig:Walkthrough}.3). The \textit{Consultant} explains that Output I is triggered when Input B detects continuous foot pressure, but Input A does not detect hand presence for 30 seconds. This ensures the yoga pad provides assistance during the Triangle Pose when hand support is needed. After Emily confirms the logic, the \textit{Consultant} informs her that the design definition is complete and asks if she wishes to finalize the design or make adjustments.


\textbf{Preview and interact with the generated device} (Fig. \ref{fig:Walkthrough}.4). After reviewing the design description, Emily clicks the “Result” button to view the implementation solution. This interface includes the DG, logic, inspection report, 3D models of the hardware, and detailed suggestions for the I/O Module, such as the fabrication pattern of the Output module. Emily can interact with the each input module to change its state and preview the corresponding output changes (Fig. \ref{fig:Walkthrough}.4.1). With the model and description, Emily is ready to assemble the fluidic circuit, fabricate inflatable airbags (Fig. \ref{fig:Walkthrough}.4.2), and integrate the FCI into a real-world application.

\section{Performance} 
In this section, we evaluate the GDT’s performance with a particular focus on two key aspects: its ability to propose DGs and its capacity to realize them. During the experiments, we provided only basic prompts, such as asking the tool to propose DGs, requiring it to independently complete each component’s design, and requesting a check after each design.

\subsection{Performance in Proposing Design Goals}
\begin{figure*}
    \centering
    \includegraphics[width=\linewidth]{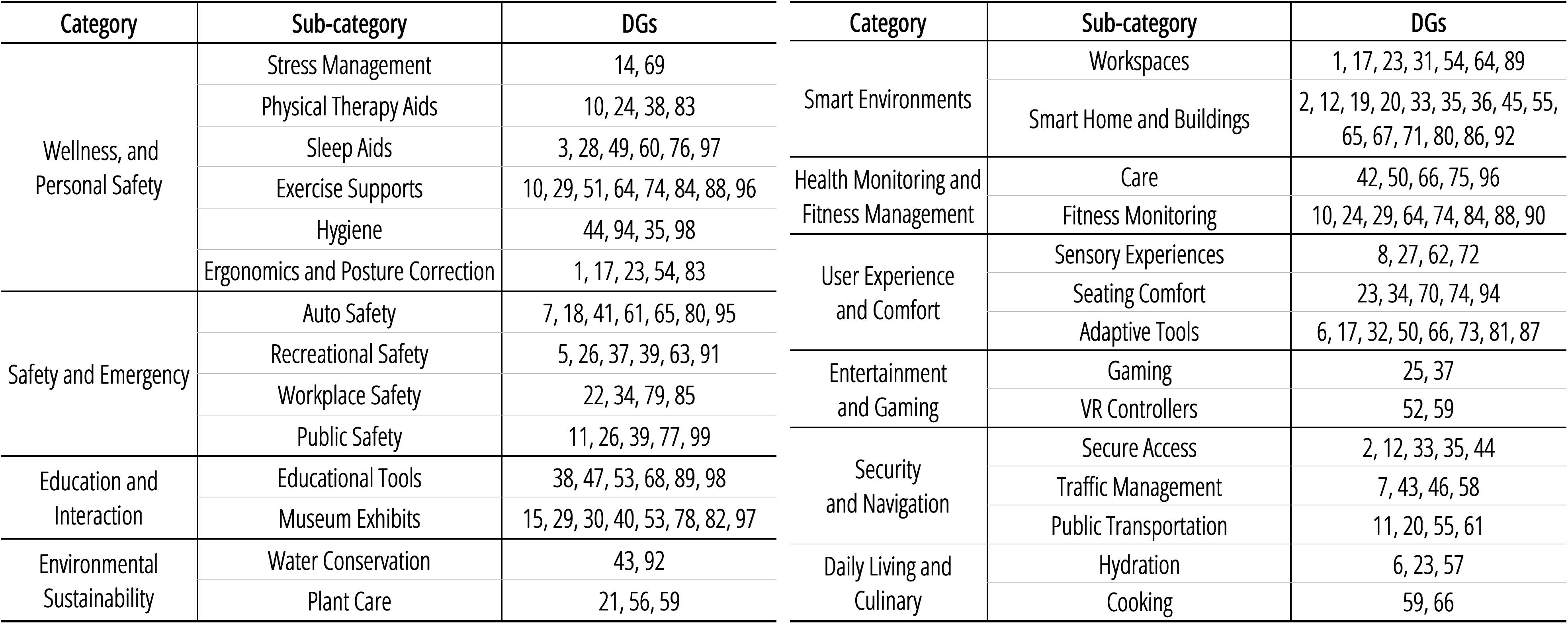}
    \caption{Categorization of the one hundred design goals proposed by the GDT, based on their application scenarios.}
    \label{fig:diversity}
    \Description{}
\end{figure*}

\begin{figure}[b]
\vspace{-0.5cm}
    \centering
    \includegraphics[width=0.5\linewidth]{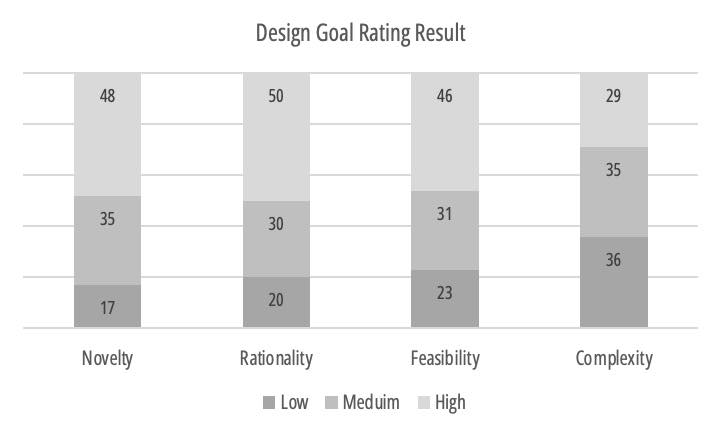}
    \caption{Rating result of the one hundred design goals.}
    \label{fig:rating}
    \Description{}
\end{figure}


We designed the \textit{Consultant} to ensure that when proposing DGs, it includes the function, necessary inputs and feedback. Overall, the GDT consistently provided DGs that met these criteria. It was tasked with generating 100 DGs, delivering 10 at a time over 10 sessions. The full list of DGs is available in the Supplementary Material. First, we evaluated the variety of these goals, categorizing them by application scenarios (Fig. \ref{fig:diversity}). This categorization covers ten major categories and twenty-four subcategories, demonstrating significant diversity. Next, we evaluated the DGs based on four criteria:

\textbf{Novelty}: This criterion evaluates the uniqueness and innovation of the application scenario, focusing on how original the design is compared to existing literature and other DGs proposed by the GDT. For example, a design like “…an encrypted door latch responding to a specific sequence of steps with a particular frequency…(DG2)” lacks novelty due to its similarity to previous works. In contrast, “…promoting an interactive learning environment by detecting edge transitions in pressure on a classroom floor…(DG89)” stands out as novel, being both unique and without similar applications in the experts’ knowledge or other proposed DGs.

\textbf{Rationality}: This criterion assesses the logical foundation and appropriateness of the design’s function within its intended scenario, ensuring the feedback or interaction is coherent and suited for its purpose. For example, “…ensure driver attentiveness by detecting the frequency of steering wheel manipulation…(DG16)” is considered irrational due to a weak correlation between the elements. In contrast, “…encourage hydration during exercise by detecting the absence of bottle squeezing and producing an olfactory reminder…(DG49)” is considered more rational.

\textbf{Feasibility}: This criterion assesses the likelihood of the proposed interface being practically realized using the kit. For instance, “…sensing the rising and falling edges of ambient noise pressure on sound-absorbing panels…(DG79)” is deemed infeasible due to the lack of technology capable of discerning noise levels.

\textbf{Complexity}: This criterion measures the complexity of the DGs, considering the number of inputs, outputs, and computational logic. Experts assess whether a) multiple inputs or outputs are required, and b) the computation demands more than simple one-to-one logic or cascading operators. A design is classified as low complexity if neither condition is met, medium complexity if one condition is met, and high complexity if both are met.

The scoring results for the 100 DGs are shown in Fig. \ref{fig:rating}. Medium or high novelty was observed in 83\% of the DGs, while complexity was evenly distributed across the three levels. Rationality and Feasibility, critical for realizing DGs, showed that about 80\% scored medium or higher in both. For DGs scoring low in Rationality or Feasibility, we conducted additional tests by feeding them back to the GDT with a prompt explaining these criteria, asking for optimization. After re-evaluation by experts, the number of DGs scoring low dropped from 20 to 7 in Rationality and from 23 to 4 in Feasibility.

\subsection{Performance in Realizing Design Goals}
From the pool of DGs with high novelty and medium or higher Rationality/Feasibility, we selected nine DGs—three from each complexity level. The GDT was tasked with creating three designs per DG, resulting in 27 designs. The selected outcomes are illustrated in Fig. \ref{fig:complexity1}, \ref{fig:complexity2}, and \ref{fig:complexity3}. Due to space limitations, we detailed 9 of these designs. We evaluated their accuracy and highlighted both strengths and challenges encountered during the design process.

\subsubsection{Accuracy} Experts evaluated design accuracy based on whether the FCI design result could achieve the DGs. Accuracy was higher in designs with low to medium complexity (Fig. \ref{fig:complexity1}, \ref{fig:complexity2}) but declined with high-complexity designs (Fig. \ref{fig:complexity3}).

\subsubsection{Design Process} We summarized the pros and cons observed in the design process, categorized by the steps following the establishment of a DG.

\begin{figure*}
    \centering
    \includegraphics[width=1\linewidth]{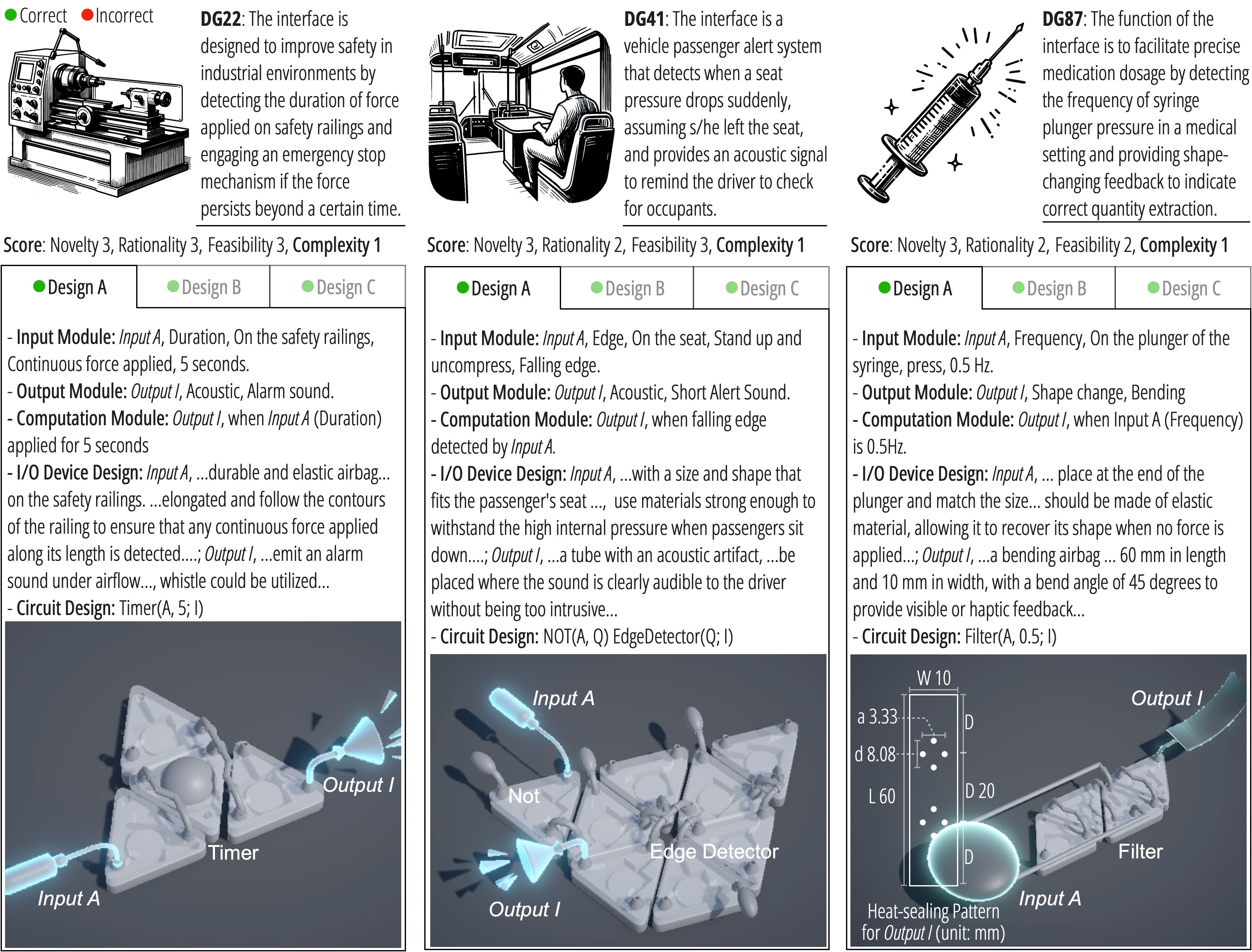}
    \caption{Selected design outcomes for three low-complexity Design Goals (DGs). For DG87-A, the LLM agents opted for a bending shape-changing output, a predefined shape supported by a corresponding inverse design algorithm. Consequently, the GDT provided a heat-sealing pattern, with its dimensions calculated according to the bending strip dimensions suggested by the agent.}
    \label{fig:complexity1}
    \Description{}
\end{figure*}

\begin{figure*}
    \centering
    \includegraphics[width=1\linewidth]{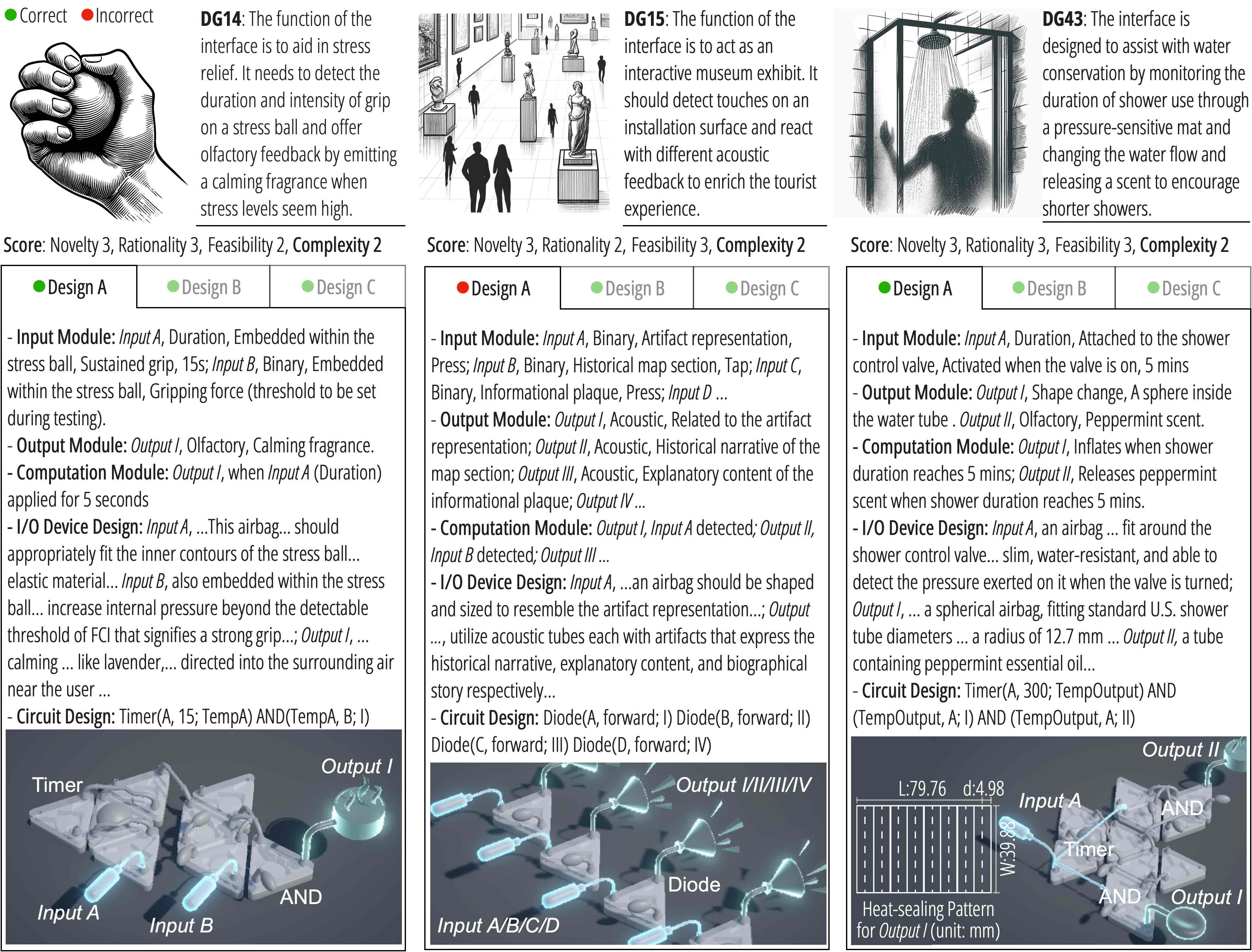}
    \caption{Selected design outcomes for three medium-complexity DGs. For DG15-A, The circuit was deemed faulty because the input airbag can only provide a signal and cannot supply a continuous airflow. The diode should be replaced with two NOT gates.}
    \label{fig:complexity2}
    \Description{}
\end{figure*}

\begin{figure*}
    \centering
    \includegraphics[width=1\linewidth]{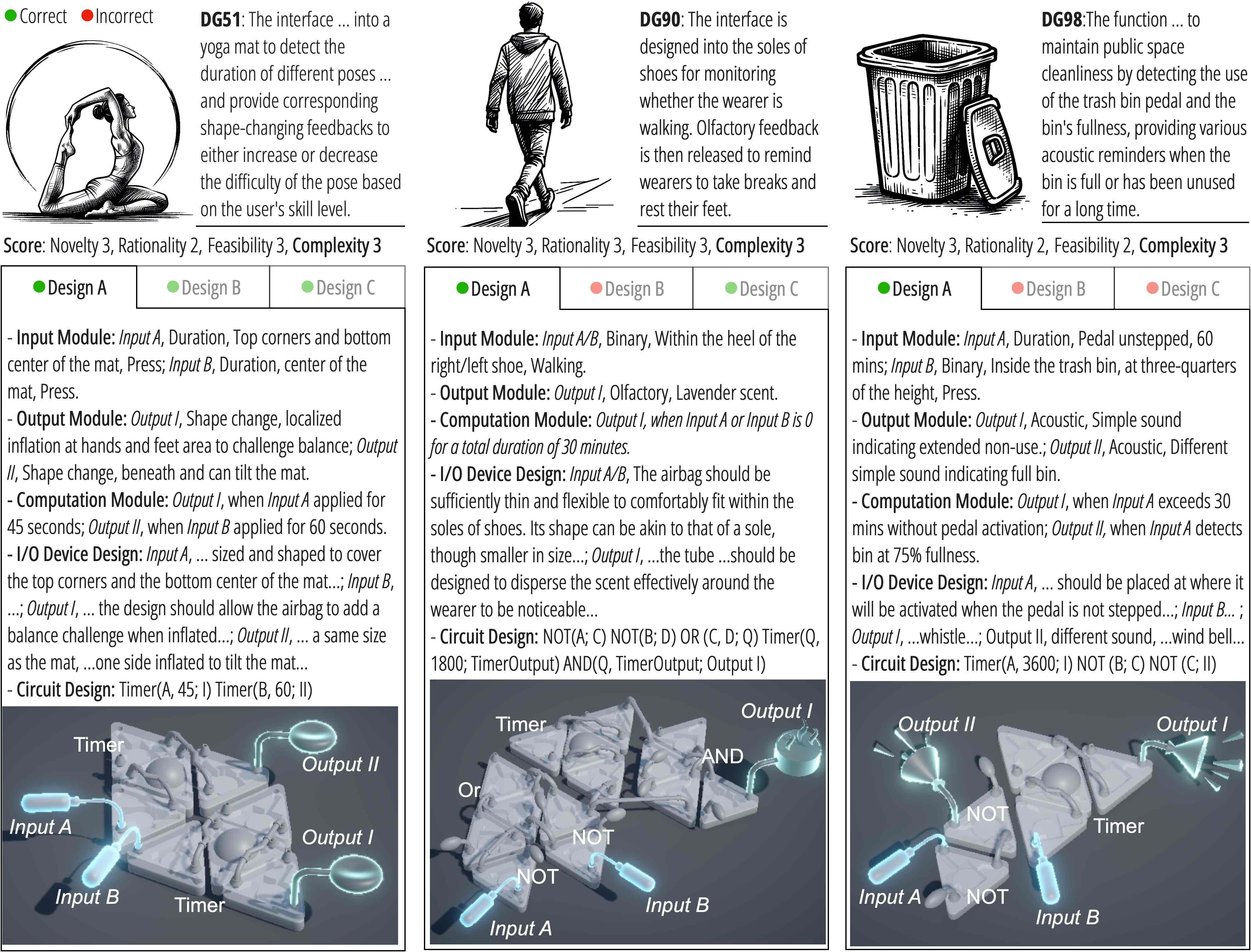}
    \caption{Selected design outcomes for three high-complexity DGs. DG90-B was unsuccessful in proposing a logical rationale to distinguish between walking and standing. DG98-B did not propose a feasible solution for detecting when the trash pedal is not stepped on. Furthermore, both DG98-B and DG98-C did not offer a viable location for placing the Input B airbag to detect when it is 75\% full.}
    \label{fig:complexity3}
    \Description{}
\end{figure*}

\textbf{I. Defining Input/Output Module.} 
Overall, \textit{the agent effectively determines the appropriate number and form of inputs/outputs (I/O) based on the DG}. For example, “Given the design goal of facilitating precise medication dosage…, it’s appropriate to have a single output module corresponding to the single input… (Fig. \ref{fig:complexity1}, DG87).” When the DG lacks explicit details on I/O, the agent infers reasonable decisions from the context, such as “…to distinguish between standing still and walking, we could use the Frequency attribute to detect pressure changes in the sole (Fig. \ref{fig:complexity3}, DG90).” This DG does not specify what to detect for determining walking, yet the agent thoughtfully decides on a rational input design. Similarly, for a yoga mat design, “ we’ll define inputs based on common poses and the areas of the mat detecting pressure during those poses. For a basic yoga session, we can plan for two inputs: Input A for Downward Dog, detecting pressure on the hands and feet, and Input B for Seated Forward Bend, detecting pressure at the center of the mat (sit bones). The input attribute would be Duration, measuring how long the pose is held, with the manipulation being the sustained press of the body against the mat. (Fig. \ref{fig:complexity3}, DG51).” The agent explains the number of inputs, their Location, and Manipulation attributes. \textit{When an I/O form in DGs seems impractical, the agent can also proactively adjust}, as in “In an industrial setting, haptic or acoustic feedback may be more suitable than shape-changing mechanisms, which could interrupt operations (Fig. \ref{fig:complexity1}, DG22).”

We also instruct the \textit{Consultant} to \textit{determine values for certain Input Attributes}, such as Frequency and Duration. The agent often provides reasonable suggestions, like “in a medical setting where precise dosage is critical, a lower syringe plunger pressure frequency, around 0.5 Hz, would allow for careful measurement… (Fig. \ref{fig:complexity1}, DG87)” or “…tracks the duration that the shower mat is under pressure, with a default threshold of 5 minutes, aligning with environmental recommendations. (Fig. \ref{fig:complexity2}, DG43).” Similarly, for Output Feedback, the agent suggests appropriate feedback types, such as “The scent should have calming properties, like lavender or chamomile… (Fig. \ref{fig:complexity2}, DG14).”

\textit{While the agent often provides appropriate values initially, it doesn’t always get them right on the first attempt. However, prompt- ing a self-check can usually correct such errors}. For example, the frequency was first set at 5 Hz, then adjusted to 0.5 Hz after a self-check (Fig. \ref{fig:complexity1}, DG87). Occasionally, the agent overlooks the limitations of the fluidic computation kit, particularly with acoustic feedback, as seen in "Output II, Acoustic, Historical narrative of the map section (Fig. \ref{fig:complexity2}, DG15)." This oversight could be due to the descriptions of available sound feedback forms in the knowledge base not being sufficiently clear to the agent.

\textbf{II. Defining Computation Module.} 
\textit{For DGs with medium complexity and below, the agent generally proposes computation logic that aligns with the DG}. For example, "Given our design goal…, the output should be triggered when the input detects a falling edge, indicating a sudden drop in seat pressure. Thus, the condition for activating the acoustic feedback is when a falling edge is detected by Input A (Fig. \ref{fig:complexity1}, DG41)."

However, \textit{for DGs with high complexity, the agent begins to encounter errors more frequently}. For example, "if input A detects that the Downward Dog pose is held for less than 30 seconds, it suggests the user is a beginner, and Output I will inflate for support. If held for more than 60 seconds, it indicates a higher skill level, and Output I will deflate to increase difficulty… (Fig. \ref{fig:complexity3}, DG51)." The flaw here is assuming all users are beginners within the first 30 seconds, regardless of skill level. \textit{Typically, allowing the agent to perform a self-check can correct such errors}: "if input A detects the Downward Dog pose for more than 45 seconds, it indicates a higher skill level, and Output I will deflate to increase difficulty (Fig. \ref{fig:complexity3}, DG51)."

\textbf{III. I/O Module Design.}
In terms of the \textit{input module, the corresponding agent typically provides accurate qualitative information}, including the design (size, shape, material) of the airbag as an input device and its placement. For example, "The airbag should be sufficiently thin and flexible to comfortably fit within the soles of shoes, with a shape similar to a sole but smaller in size. (Fig. \ref{fig:complexity3}, DG90)," "…the airbag should be made of elastic material, allowing it to recover its shape when no force is applied, catering to the 'Frequency' attribute's need for resiliency… (Fig. \ref{fig:complexity1}, DG87)," and "…design an airbag sized and shaped to cover the top corners and bottom center of the yoga mat… (DG51)." However, we observed a \textit{weakness in understanding spatial distribution}, as the agent does not actively avoid interference between multiple input devices in the same space, as shown by "For Input A, …This airbag… should appropriately fit the inner contours of the stress ball… For Input B, another airbag embedded within the stress ball is needed to detect binary gripping force… (Fig. \ref{fig:complexity1}, DG14)."

Regarding the \textit{output module, the agent generally provides suitable qualitative suggestions as well}. For instance, "For Output I, a tube designed to emit an alarm sound under airflow is needed. A pneumatic whistle could be utilized… (Fig. \ref{fig:complexity1}, DG22)" and "…for olfactory feedback, a tube imbued with a refreshing scent, filled with essential oils known for their invigorating properties, should release the scent into the shower area (Fig. \ref{fig:complexity2}, DG43)." For "Shape-changing Feedback" that matches \textit{preset shapes in the knowledge base, we tasked the agent with quantifying relevant geometry based on the context}. The agent performs well in these tasks, as shown by "Since Output I is an airbag that inflates inside the water tube to partially obstruct flow after 5 minutes of shower use, its design must allow for expansion to effectively regulate water flow. A spherical airbag, fitting standard U.S. shower tube diameters, with a radius of 12.7 mm (0.5 inches), is optimal, significantly reducing but not stopping the flow (Fig. \ref{fig:complexity3}, DG43)" and "Given the unspecified syringe type, for common usage, a bending airbag could measure 60 mm in length and 10 mm in width, with a bend angle of 45 degrees, attachable to the syringe's label area (Fig. \ref{fig:complexity1}, DG87)." 

\textit{Inverse design for obtaining the airbag's heat-seal pattern is consistently accurate}, as seen in "L: 79.76 mm, W: 39.88 mm, d=4.98 (Fig. \ref{fig:complexity3}, DG43)" and "L: 60 mm, W: 10 mm, a: 3.33 mm, d: 8.08 mm, D: 20 mm, n: 2 (Fig. \ref{fig:complexity1}, DG87)." However, we noted a \textit{shortfall in accurately identifying more intricate spatial relationships and physical interactions}. For example, DG98 (Fig. \ref{fig:complexity3}) demonstrated the agent's difficulty in determining the appropriate placement for the input airbag, which was crucial for detecting bin fullness.

\textbf{IV. Computation Module Design.}
This task is collaboratively completed by the \textit{Logic Designer}, \textit{Circuit Engineer}, and \textit{Inspector}. \textit{The Logic Designer accurately transcribes and standardizes the DG and module design information into a "truth table" and demonstrates the ability to refine computation logic}. For example, "If the duration of A or B being 0 is < 30 minutes, then Output I = 0; If the total duration of A or B being 0 is >= 30 minutes and A or B = 0, then Output I = 1 (Fig. \ref{fig:complexity3}, DG90)," where the \textit{Logic Designer} introduces additional AND logic to ensure the output is activated only when the walking duration exceeds the specified time and the subject is still walking.

\textit{The Circuit Engineer performs well in designing circuits for DGs} with medium or lower complexity but may introduce errors in high-complexity DGs. For example, "\textit{OR (A, B; Q) Timer(Q, 1800; TimerOutput) AND(Q, TimerOutput; Output I)} (Fig. \ref{fig:complexity3}, DG90)" contains a mistake where Input A is not inverted, causing incorrect detection when A is pressed. In such cases, \textit{the Inspector identifies the errors and suggests corrections}, "The truth table specifications are not fully met by the current circuit description… an inversion of Input A and Input B's condition is required…, 'circuit': NOT(A; C) NOT(B; D) OR (C, D; Q) Timer(Q, 1800; TimerOutput) AND(Q, TimerOutput; Output I) (Fig. \ref{fig:complexity3}, DG90)."

\section{Implementation} 
In this section, we detail the implementation specifics of our GDT.

\subsection{The Technical Architecture}

\begin{figure}[t]
    \centering
    \includegraphics[width=0.7\linewidth]{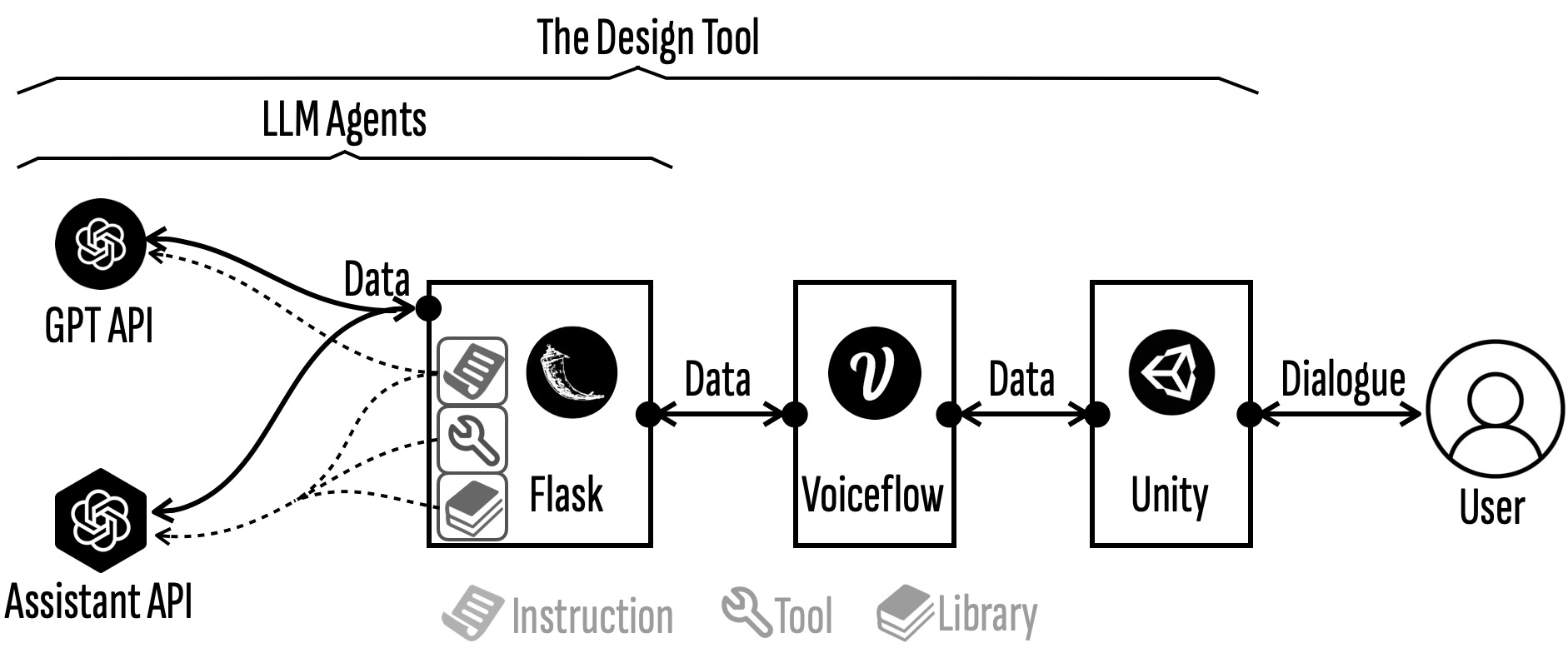}
    \caption{The technical architecture of GDT.}
    \label{Fig:tool_framework}
    \Description{}
\end{figure}

The architecture, illustrated in Fig. \ref{Fig:tool_framework}, integrates OpenAI APIs (GPT4-preview), Voiceflow, Flask, and Unity. Unity handles the front-end, while Voiceflow, Flask, and OpenAI APIs manage the back-end, enabling the collaborative workflow of LLM agents.

Agents are powered by Flask and OpenAI APIs. Flask, a lightweight Python web framework, facilitates API development, request handling, data post-processing, and serves as the integration layer. Resources such as instructions, libraries, and tools are hosted on Flask and relayed to OpenAI APIs during agent initialization.

We use two distinct OpenAI APIs: GPT and Assistant. 
The Assistant API differs by enabling tool use, maintaining contextual memory, and consulting a knowledge base (the Library), though it incurs higher token usage. The choice between GPT and Assistant depends on task complexity. For basic tasks like those of the \textit{Logic Designer} and \textit{Inspector}, the GPT API is ideal for its simplicity. For more complex roles, such as the \textit{Consultant}, \textit{I/O Designer}, and \textit{Circuit Engineer}, the Assistant API's advanced features are more advantageous.

Flask assigns a specific port to each agent, acting as its unique "address" within the public network for easy access by Voiceflow. These dedicated ports allow Voiceflow to facilitate communication between agents, enabling a smooth collaborative workflow. Voiceflow orchestrates agent interactions, while Unity (Version 2022.3.15) handles the front-end interface, monitoring exchanges via the Voiceflow API to ensure seamless user interaction. Along with managing conversations and displaying text-based results, Unity also integrates algorithms for visualizing and simulating the FCI.

\subsection{Configure LLM Agents}

\begin{figure}[b]
    \centering
    \includegraphics[width=0.4\linewidth]{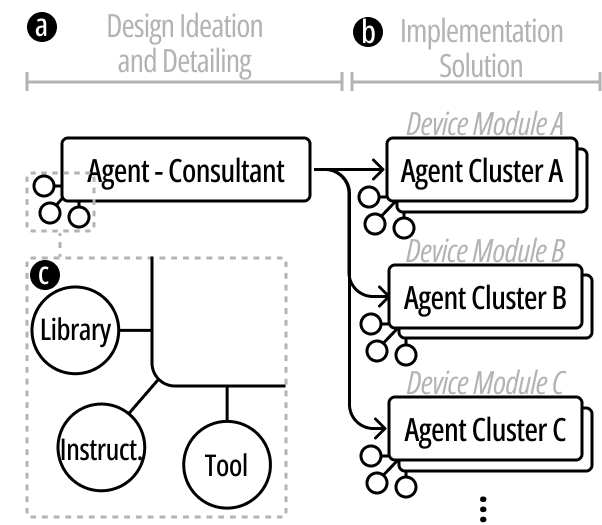}
    \caption{A potential general agent collaboration framework. Our GDT consists the \textit{Consultant} and two agent clusters (one for solving the computation module design, one for the I/O Modules). }
    \label{fig:agent_framework}
    \Description{}
\end{figure}

\subsubsection{\textbf{Task Allocation for Multi-Agents}}
Assigning agents to focus on narrow, specific tasks typically leads to better outcomes \cite{li2024camel, li2023metaagents}. In the context of designing novel devices like FCIs, we organize and delegate tasks based on the following considerations:

Initially, we divide the design process into two main phases: Ideation \& Detailing and Implementation Solution. The Ideation \& Detailing phase requires a comprehensive qualitative understanding of the design space for the new technology (in this case, the FCI) and a macroscopic view of the design. This sub-task is assigned to a single agent (\textit{Consultant}), which is guided to gather or propose sufficient design details (Fig. \ref{fig:agent_framework}.a).

The Implementation Solution phase focuses on the specific design of sub-modules within the device, requiring specialized, in-depth knowledge, which can be assigned to different clusters of agents (Fig. \ref{fig:agent_framework}.b). For the FCI, these modules are divided into Input, Output, and Computation. However, we found that some knowledge required for designing I/O modules is difficult to effectively communicate to LLM agents through natural language, and LLM agents tend to struggle with tasks involving structure, geometry, and space. As a result, we adjusted the expectations for these modules to primarily generate qualitative suggestions, simplifying the task for a single agent (\textit{I/O Designer}). 

For the Computation Module, since LLM has a foundation in electronic circuit design, we set the expectation to produce directly executable solutions. Following standard logic design practices \cite{dessouky2019hardfails, blocklove2023chip}, we divide the task into Logic Analysis (\textit{Logic Designer}), Circuit Generation (\textit{Circuit Engineer}), and Circuit Inspection (\textit{Inspector}), assigning these tasks to an agent cluster to optimize performance.

\subsubsection{\textbf{Qualifying Agents for Tasks}}\label{sec:qualifyingtheagent}
We primarily qualify agents for their assigned tasks through three resources: the Libraries, the Instruction, and the Tool (Fig. \ref{fig:agent_framework}.c).

\textbf{I. Instructions} are carefully crafted prompts that articulate the objectives and outline the expected method for the agent's response. Detailed Instructions for each agent can be found in Appendix \ref{sec:instructions}. Proper organization of these Instructions is crucial to enhancing the agent's effectiveness. To achieve this, we structure the Instructions into distinct sections, simplifying comprehension for the agents and easing maintenance for developers. Generally, the sections include:

\begin{itemize}
    \item \textit{Role} delineates the fundamental identity and capabilities of each agent, providing a clear directive like "you are an electronic circuit engineer, your goal is …".     
    
    \item \textit{Input Interpretation} enhances an agent's comprehension of information received from others, ensuring accurate understanding and response.
    
    \item \textit{Resources} serve as the agents' memory, embedding task-specific knowledge relevant to their roles, such as "I/O module design principles" for the \textit{I/O Designer}. Another example is the Operator Description Language, created for the \textit{Circuit Engineer} and \textit{Inspector} (Appendix \ref{sec:CircuitEngineer}, \ref{sec:Inspector}), which efficiently conveys the assembly of FCI's basic computational units. By digitizing analog operators like filters and timers, it reduces the agents' cognitive load during circuit design. Information not immediately needed is stored in the Library and accessed by agents only when necessary, preventing the Instructions from becoming overwhelming or distracting.

    \item \textit{Tools} convey to the agent the suite of functions it can utilize, presented more as an accessible list than detailed within the Instructions.

    \item \textit{Workflow} provides a structured approach for task execution, guiding agents through complex activities. For instance, we offer a reference dialogue logic for the \textit{Consultant} to ensure a comprehensive collection of design details.

    \item \textit{Attention} section directs the agent's focus to critical elements requiring extra scrutiny. For example, the \textit{I/O Designer} is reminded to use specific functions for calculating the dimensions of the heat-sealing pattern when certain shapes are specified for the output module.

    \item \textit{Output Requirement} delineates the anticipated format, categories, and constraints for the agent's outputs. Setting these standards facilitates smoother communication among agents.
    
\end{itemize}

\textbf{II. Library} (Appendix \ref{sec:Library}).
In our context, the Library stores comprehensive knowledge about the FCI's design space. The content, distilled and abstracted from academic papers \cite{lu_fluidic_2023}, retains core information necessary for agents to understand FCI. It is meticulously organized and tagged to facilitate easy retrieval. The Library primarily covers the basic concepts and framework of FCI, along with detailed explanations of the input, output, and computation modules. Additionally, various examples for each module are included to enhance understanding.

\textbf{III. Tools} (Appendix \ref{sec:Tools})
Tools primarily consist of functions that assist agents in executing specific tasks. For instance, the \textit{Consultant} can use functions to save basic information to local JSON files after completing a module design, facilitating a smooth transition to the next design phase. Another set of tools includes inverse design functions for the output module when shapes like spheres, cylinders, boxes, folds, or bends are selected \cite{teng_pupop_2018, yang2024snapinflatables} (Appendix \ref{sec:inversedesign}). The \textit{I/O Designer} invokes the relevant function as needed.

\subsubsection{\textbf{Facilitating Agents Collaboration}}
To facilitate collaboration among agents, we implement \textit{standardized information formats} for communication and ensure the agents \textit{receive necessary information only}. 

\textbf{I. Standardized Information Formats}.
The use of well-defined structured information, such as Standardized Operating Procedures (SOPs) \cite{gourevitch2008standard}, has been shown to improve task consistency and accuracy \cite{qian2023communicative, hong2023metagpt}. Similarly, we require each agent to output information in a standardized JSON format to enhance cooperation and streamline information extraction. For example, the output from the \textit{Circuit Engineer} should follow a format like \textit{\{'circuit': 'Filter(input, 3; output)', 'description': 'The Filter operator is used to ...'\}}. This standardization enables circuit details to be extracted for visualization in Unity and evaluated by the \textit{Inspector}.

\textbf{II. Receive Necessary Information Only} 
To avoid information overload and improve the efficiency of our agents, we ensure they receive only the information critical to their specific roles. Excessive, irrelevant data can dilute focus and lead to outputs that deviate from established standards. This decline in performance is partly due to the reliance of LLMs on the Transformer architecture, which operates on attention mechanisms \cite{vaswani2017attention}. For example, we observed that the Circuit Engineer's proficiency in designing efficient circuits decreased when it was exposed to non-essential information.

\subsection{Circuit Visualization and Simulation}
We implemented the core algorithm for circuit construction using C\# in Unity. It is mainly divided into three parts: FCI operator construction, layout wiring, and simulation.

\begin{figure}[b]
    \centering
    \includegraphics[width=0.6\linewidth]{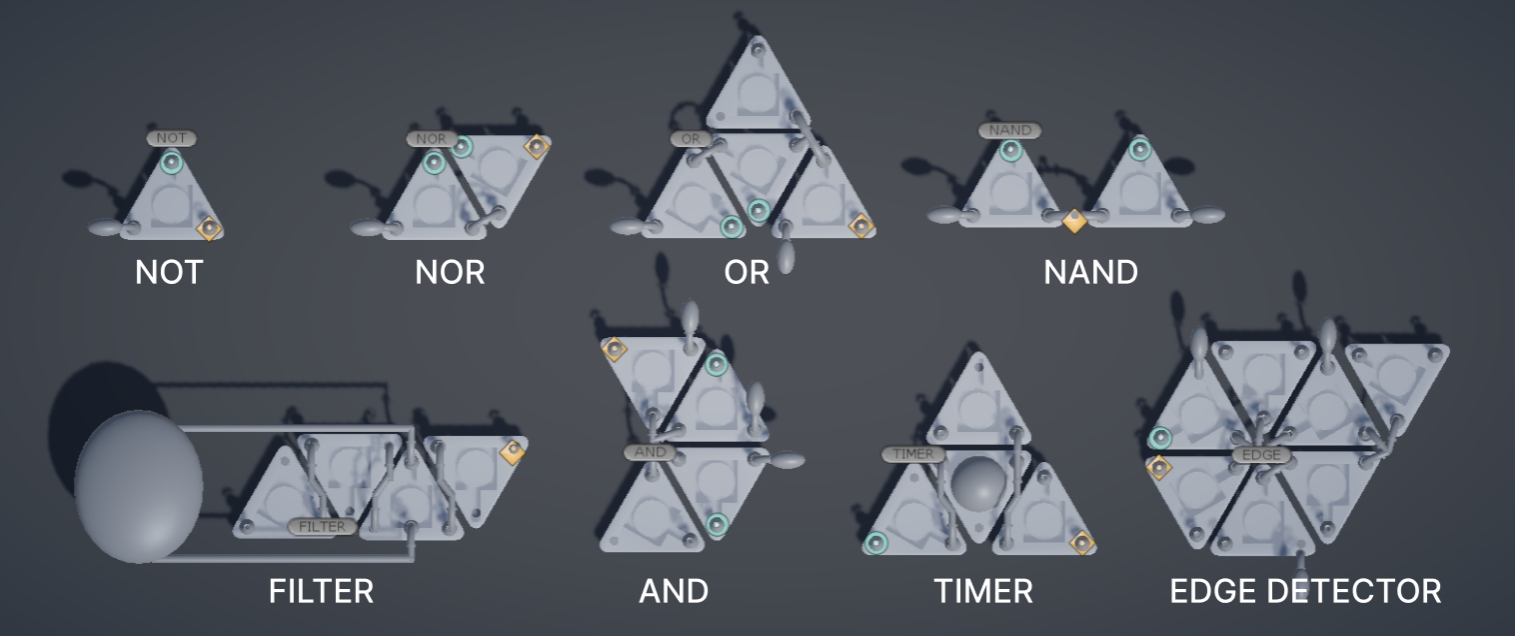}
    \caption{The model of the FCI Operator and its corresponding input/output ports.}
    \label{fig:gate}
    \Description{}
\end{figure}

\subsubsection{FCI Operator}
We created operator models in Blender and defined their data structures upon importing them into Unity (Fig. \ref{fig:gate}). Each operator has an associated template, including cell/port positions, input count, and a simulation function. Circuit parsing starts with the final JSON format generated during the design phase, which details all logic operators and their connections. The parsing function converts this description into an internal network structure, instantiating the circuit. This process assigns a unique node to each logic operator, records connections, and generates network ports for each operator.

\subsubsection{Layout Wiring}
We employed a simulated annealing algorithm to optimize the physical layout, aiming to minimize the area and wire crossings. Starting from a random initial layout, the system explores the layout space by randomly adjusting the positions and rotation angles of the logic operators. After each adjustment, a target function evaluates the layout's quality based on operator overlap, wire length, and the total area occupied. This process results in a compact and optimized circuit design.

\subsubsection{Simulation}
The simulation process begins with the user's click events on the input model. The system navigates through each logic operator in the circuit network, computing the output based on predefined operator logical functions. A dictionary stores and retrieves the signal states of each operators. Additionally, a list is created to archive circuit components and signal propagation pathways, ensuring signal processing follows the circuit's design sequence. To handle the asynchronous nature of signal propagation, a specialized queue ensures accurate signal transmission order across operators, simulating dynamic changes. The output model will then display an animation upon activation.

\section{Reflection, Discussion \& Future Work}
In this section, we reflect on the advantages, limitations and scalability of LLM agents in the context of enhancing design tools for novel devices. 

\subsection{Breadth of Knowledge and Divergent Thinking}
LLM agents offer a vast knowledge base and a unique ability to generate diverse ideas, complementing human creativity during the design ideation phase. This is especially beneficial for generating innovative solutions across various scenarios, as LLM agents can quickly access and synthesize information from a wide range of topics, broadening perspectives and sparking creative thinking.

Furthermore, in the DG detailing phase, the synergy between human intuition and LLM's data-driven suggestions can yield remarkable results. For instance, beyond proposing specific scents for olfactory feedback in FCIs, the LLM agent can tap into its database to suggest essential oil recipes, enhancing the process with more informed options.

However, the extensive knowledge base and idea generation capacity of LLM agents can become a drawback if not properly managed. It is essential to define the design space and capabilities of the new technology for the agent clearly. Setting constraints is also crucial; for example, while FCIs can be compatible with electronic sensors or actuators, the focus should remain on non-electronic designs. Explicitly informing the agent that FCI is a non-electronic system can significantly reduce the likelihood of it suggesting inappropriate electronic components. This ensures the LLM agent’s contributions adhere to the technology.

\subsection{Specialized Knowledge and Areas of Limited Proficiency}
Designing novel devices involves specialized technical tasks. If the relevant knowledge is well-represented in the LLM's training data, the agent generally performs competently, as seen with tasks like programming \cite{hong2023metagpt,chen2023autoagents} and electronic circuit design \cite{blocklove2023chip}. However, if the knowledge is not present, the LLM’s ability to handle the task could highly depend on how effectively the new information is explained and whether it can leverage its existing capabilities. For instance, the “circuitry” in FCIs represents a novel approach to mechanical computation but shares similarities with electronic circuits, a domain covered in LLM training. After explaining the logic operators used in FCI, the agent can apply its knowledge of electronic circuitry to design the computation module for FCIs.

On the other hand, the design intricacies of I/O modules within FCIs present a challenging domain for LLM agents. This is largely due to the specialized knowledge required to understand the complex behaviors and structures of soft actuators. Currently, our tuned agent in the GDT can mainly provide qualitative guidance in these areas. While it can suggest quantitative dimensions for a few simple shapes that have been explained to it, its ability to handle tasks requiring a deep understanding of spatial geometry remains limited. This constraint applies both to introducing new geometric concepts and expecting the agent to design complex geometric shapes or mechanisms.

Efforts to overcome spatial comprehension challenges are crucial, especially since many novel device technologies rely heavily on complex spatial and mechanical knowledge. Enhancing an LLM agent's capabilities in spatial reasoning and mechanical design could involve enriching its training data of related scenarios. However, it is important to recognize that certain domains, such as 3D modeling, may inherently fall outside the LLM's optimal skill set due to its text-based knowledge and reasoning nature. In such cases, leveraging other types of generative AI specialized in 3D design might be more effective \cite{jun2023shap, shi2023mvdream}. These specialized AI systems can complement LLM agents by providing expertise in areas where LLMs face limitations, enabling a more comprehensive approach to the design of novel devices.

\subsection{Contextual Awareness}
Throughout the design task, we observed that the agent demonstrated strong contextual awareness and cognitive capabilities, enabling a generative design process beyond the capabilities of conventional design tools. We identified three tiers of contextual awareness exhibited by the agents:
\begin{itemize}
    \item \textbf{Understanding the Design Space:}
    The agent, when properly tuned, demonstrates a solid awareness of the FCI’s design space. This allows it not only to grasp the broader technological landscape but also to identify and propose appropriate DGs.
    \item \textbf{Reasonably Detailing Post-DG Establishment:}
    Once a DG is established, the agent can propose sensible design details and correct errors based on context. For example, it can suggest representative yoga poses along with corresponding input detection methods and recommend feasible parameters, such as frequency and duration, based on the intended detection target or action.
    \item \textbf{Strategic Technical Implementation:}
    When determining technical implementation, the agent can propose practical qualitative and even quantitative solutions based on the DG and design details. Examples include advising on the shape and material requirements for input airbags or suggesting dimensions for shape-changing output airbags tailored to the usage scenario.
\end{itemize}
We believe that the contextual awareness of LLM agents could enhance the process of designing novel devices. Our hypothesis suggests that such awareness not only streamlines the ideation and development phases but also promotes the creation of solutions that blend innovation with practicality. We vision that the agent’s contextual awareness could become a key factor in navigating the complexities of novel device design. If LLM agents can demonstrate the ability to understand, detail, and implement design strategies with precision, they could become invaluable assets in the evolving landscape of novel device creation.


\subsection{Configuring and Debugging Agents}
Configuring and debugging the agent is a central task when implementing the GDT. A key process is carefully crafting the Library and Instruction. Theoretically, the configuration process is straightforward, making it accessible even to those without extensive programming experience. Both the Library and Instructions are primarily written in natural language and can be structured using our organizational template (section \ref{sec:qualifyingtheagent}). However, practical implementation presents challenges. Developing the Library requires distilling complex technological information while writing Instructions tests the ability to communicate effectively and succinctly to the agent. Additionally, natural language is not always the most efficient method for conveying large amounts of information, differing significantly from the highly abstract, structured, and standardized nature of coding.

Furthermore, the Library and Instructions serve more as soft constraints \cite{nam2024using}, with the agent acting based on its own "understanding and reasoning." For instance, for the \textit{Consultant} agent, we defined a dialogue inquiry flow, but it doesn't necessarily follow this process strictly and can switch between different sections as needed. Developers should recognize this dynamic and avoid trying to force the agent into specific actions. Instead, they should embrace and leverage the agent's autonomy, trusting its capabilities and allowing it some freedom in task completion. For example, the agent can be informed that it has the discretion to independently assess the collection of design details and guide users through completing various sections.

The process of debugging agents differs significantly from code debugging. In agent debugging, identifying problems can be less straightforward, as issues are often not specifically pinpointed to lines of faulty code \cite{liu2024towards}. Instead, developers must observe the agent's behaviors, gradually adjusting and refining the Library and Instructions. For example, if the agent consistently overlooks certain device functions, additional relevant knowledge may need to be added. If the agent's designs exceed the device's capabilities, new restrictions should be implemented. If the agent misinterprets a design element, it's important to review whether unclear expressions caused the confusion.

In summary, configuring and debugging the agent is an iterative, optimization-focused process, similar to writing and refining a manual. By developing efficient knowledge libraries, providing clear operational instructions, and carefully debugging agent behavior, the performance and applicability of the GDT can be improved.

\subsection{Scalability of the Approach} 
We believe that the approach used in this research, particularly the task decomposition and agent allocation framework, could be effectively applied to \textbf{other mechanical computation devices}.
For example, Digital Mechanical Metamaterials (DMM) \cite{DMM}, which involve 3D-printed devices with mechanical computation, could use a similar approach. A \textit{Consultant} agent could assist with ideation and detailing, but since the design space isn't fully outlined in the paper, it would need to be summarized to explain DMM’s capabilities to the agent—covering key concepts, input/output functions, and limitations.
For technical implementation, the device could be divided into two parts: mechanical circuitry and metamaterials \cite{metamaterial}, each handled by different agents. The mechanical circuitry design could leverage the agent’s expertise in electronic circuit design, similar to FCI-GDT, while noting key differences like DMM’s binary, bistable configurations that don’t need a constant power supply. Clear analogies between DMM’s mechanical and traditional electronic circuits would help the agent adapt. The metamaterials part could be managed by another (group of) agent, offering general suggestions on dimensions, ergonomics, and shape, even if it can’t provide precise quantitative details.
For fabrication, traditional part of design tools could handle structure visualization and circuit fabrication file generation, as shown in the DMM paper. 

For \textbf{other type of novel devices}. The applicabiliy of our approach is possible but needs furtuther study and practice.
For example, LineForm \cite{lineform}, an actuated curve interface similar to serpentine robotics, has a well-defined design space outlined in the paper, which could help a \textit{Consultant} agent understand the technology. The main task is programming the device, which could be handled by a group of agents.
First, the \textit{Consultant} agent would describe the target shape for the design scenario by identifying key geometric parameters and explaining how changes affect the shape. Another agent could then program the servos based on these parameters to control the curve and achieve the desired shape.
If using an inverse design algorithm is more efficient, the agent’s role could shift to determining the shape parameters and feeding them into the algorithm, which would calculate the adjustments needed to program the device accordingly.

Another example is FoamSense \cite{FoamSense}, which introduces a novel soft sensor that measures deformation states like compression, bending, twisting, and shearing. The paper thoroughly outlines the design process and elements, offering a strong foundation for tuning a \textit{Consultant} agent to conceptualize and detail designs.
Given the technology’s focus on sensor development, two clusters of agents could handle this task. The first cluster, guided by the \textit{Consultant} agent’s assembly of requirements (such as manipulation type and surface/object integration), would estimate the force range, determine the sensor’s optimal shape, select materials, and more. The paper’s experimental results could help the agent provide quantitative insights for these parameters.
The second cluster of agents would focus on developing Arduino code and peripheral circuits, a task well within the capabilities of LLM agents, with the potential to generate executable outcomes.

\subsection{Human-AI Co-design Experience}
Currently, the evaluation primarily focuses on objectively assessing the tool’s performance in designing FCI with limited human intervention, using only basic and neutral prompts. To better understand the potential of our approach in enhancing the user experience, especially in terms of how increased human involvement and collaboration might affect performance, further user studies are necessary.

We conducted a pilot user study as an initial step. Six participants were involved—three self-identified as experts in mechanical computation devices and three had very limited or no knowledge. The study began with a 15-minute walkthrough demonstrating how to use the tool. Participants were then tasked with designing two to three FCI devices within 45 minutes, followed by a short interview to gather their feedback. Here are some interesting preliminary findings:

Expert participants found the GDT inspiring in helping to formulate design applications. They adopted a more collaborative approach with the AI when detailing the designs, often proposing design details themselves and then refining and iterating them with the AI. Nevertheless, they consistently noted that the GDT significantly facilitated the process by assisting with implementable solutions from the design concept. Non-expert users initially tended to rely entirely on the GDT for their first attempt but gradually began taking a more active role in the design process during subsequent attempts. One participant mentioned that it felt like more than just a design tool; it also served as a teaching aid, helping her learn and understand FCI. Even those with no prior knowledge of FCI were generally able to produce functional designs with the help of the GDT.

The GDT was also effective in refining or correcting vague or incorrect inputs from users. However, we observed that if a participant insisted on an incorrect design, the GDT would proceed with the faulty information, indicating that mechanisms to improve the tool’s robustness might be necessary. Additionally, some participants expressed uncertainty about the correctness of the results, even when the GDT confirmed them. AI, including advanced models like ChatGPT, can still make mistakes, even in tasks they typically excel at. In the future, exploring ways to boost user confidence in the designs—such as by incorporating memory mechanisms that learn from past successes or failures~\cite{zhang2024survey}, or implementing enhanced frameworks for more thorough quality checks—could be valuable areas for further development~\cite{zhang2024towards}.

Lastly, some participants mentioned that while the design outcomes were rendered in a highly informative way, the input process, which only accepts text, felt somewhat plain. The current version of the tool supports idea input solely through textual descriptions. In the future, exploring support for inputs like speech, sketches, or images could be an interesting and useful enhancement~\cite{wu2023next,hu2024bliva}.

\section{Conclusion}
This study points towards the growing need for improved design tools in the development of interactive novel devices, suggesting the use of LLM agents as a practical enhancement. Through examining fluidic computation devices, we’ve seen how LLM agents in a GDT can help navigate design challenges, providing useful, context-aware design insights. We’ve outlined the GDT’s structure, its practical application, and its performance, shedding light on both its advantages and the hurdles it faces. This exploration indicates a potential path for refining design tools, combining traditional methods with LLM capabilities for a more effective approach to device prototyping.



\bibliographystyle{ACM-Reference-Format}
\bibliography{refnew}


\begin{thebibliography}{86}


\ifx \showCODEN    \undefined \def \showCODEN     #1{\unskip}     \fi
\ifx \showDOI      \undefined \def \showDOI       #1{#1}\fi
\ifx \showISBNx    \undefined \def \showISBNx     #1{\unskip}     \fi
\ifx \showISBNxiii \undefined \def \showISBNxiii  #1{\unskip}     \fi
\ifx \showISSN     \undefined \def \showISSN      #1{\unskip}     \fi
\ifx \showLCCN     \undefined \def \showLCCN      #1{\unskip}     \fi
\ifx \shownote     \undefined \def \shownote      #1{#1}          \fi
\ifx \showarticletitle \undefined \def \showarticletitle #1{#1}   \fi
\ifx \showURL      \undefined \def \showURL       {\relax}        \fi
\providecommand\bibfield[2]{#2}
\providecommand\bibinfo[2]{#2}
\providecommand\natexlab[1]{#1}
\providecommand\showeprint[2][]{arXiv:#2}

\bibitem[aut(2023)]%
        {auto2023}
 \bibinfo{year}{2023}\natexlab{}.
\newblock \bibinfo{title}{AutoGPT}.
\newblock \bibinfo{howpublished}{\url{https://github.com/Significant-Gravitas/AutoGPT}}.
\newblock


\bibitem[bab(2023)]%
        {baby2023}
 \bibinfo{year}{2023}\natexlab{}.
\newblock \bibinfo{title}{BabyAGI}.
\newblock \bibinfo{howpublished}{\url{https://github.com/yoheinakajima/babyagi}}.
\newblock


\bibitem[Achiam et~al\mbox{.}(2023)]%
        {achiam2023gpt}
\bibfield{author}{\bibinfo{person}{Josh Achiam}, \bibinfo{person}{Steven Adler}, \bibinfo{person}{Sandhini Agarwal}, \bibinfo{person}{Lama Ahmad}, \bibinfo{person}{Ilge Akkaya}, \bibinfo{person}{Florencia~Leoni Aleman}, \bibinfo{person}{Diogo Almeida}, \bibinfo{person}{Janko Altenschmidt}, \bibinfo{person}{Sam Altman}, \bibinfo{person}{Shyamal Anadkat}, {et~al\mbox{.}}} \bibinfo{year}{2023}\natexlab{}.
\newblock \showarticletitle{Gpt-4 technical report}.
\newblock \bibinfo{journal}{\emph{arXiv preprint arXiv:2303.08774}} (\bibinfo{year}{2023}).
\newblock


\bibitem[Adleman(1994)]%
        {1.4}
\bibfield{author}{\bibinfo{person}{Leonard~M Adleman}.} \bibinfo{year}{1994}\natexlab{}.
\newblock \showarticletitle{Molecular computation of solutions to combinatorial problems}.
\newblock \bibinfo{journal}{\emph{science}} \bibinfo{volume}{266}, \bibinfo{number}{5187} (\bibinfo{year}{1994}), \bibinfo{pages}{1021--1024}.
\newblock


\bibitem[An et~al\mbox{.}(2018)]%
        {Thermorph}
\bibfield{author}{\bibinfo{person}{Byoungkwon An}, \bibinfo{person}{Ye Tao}, \bibinfo{person}{Jianzhe Gu}, \bibinfo{person}{Tingyu Cheng}, \bibinfo{person}{Xiang~'Anthony' Chen}, \bibinfo{person}{Xiaoxiao Zhang}, \bibinfo{person}{Wei Zhao}, \bibinfo{person}{Youngwook Do}, \bibinfo{person}{Shigeo Takahashi}, \bibinfo{person}{Hsiang-Yun Wu}, \bibinfo{person}{Teng Zhang}, {and} \bibinfo{person}{Lining Yao}.} \bibinfo{year}{2018}\natexlab{}.
\newblock \showarticletitle{Thermorph: Democratizing 4D Printing of Self-Folding Materials and Interfaces} \emph{(\bibinfo{series}{CHI '18})}. \bibinfo{publisher}{Association for Computing Machinery}, \bibinfo{address}{New York, NY, USA}, \bibinfo{pages}{1–12}.
\newblock
\showISBNx{9781450356206}
\urldef\tempurl%
\url{https://doi.org/10.1145/3173574.3173834}
\showDOI{\tempurl}


\bibitem[Andreas(2022)]%
        {andreas2022language}
\bibfield{author}{\bibinfo{person}{Jacob Andreas}.} \bibinfo{year}{2022}\natexlab{}.
\newblock \showarticletitle{Language models as agent models}.
\newblock \bibinfo{journal}{\emph{arXiv preprint arXiv:2212.01681}} (\bibinfo{year}{2022}).
\newblock


\bibitem[Bhavya et~al\mbox{.}(2023)]%
        {CAM}
\bibfield{author}{\bibinfo{person}{Bhavya Bhavya}, \bibinfo{person}{Jinjun Xiong}, {and} \bibinfo{person}{Chengxiang Zhai}.} \bibinfo{year}{2023}\natexlab{}.
\newblock \showarticletitle{CAM: A Large Language Model-based Creative Analogy Mining Framework} \emph{(\bibinfo{series}{WWW '23})}. \bibinfo{publisher}{Association for Computing Machinery}, \bibinfo{address}{New York, NY, USA}, \bibinfo{numpages}{12}~pages.
\newblock
\showISBNx{9781450394161}
\urldef\tempurl%
\url{https://doi.org/10.1145/3543507.3587431}
\showDOI{\tempurl}


\bibitem[Blocklove et~al\mbox{.}(2023)]%
        {blocklove2023chip}
\bibfield{author}{\bibinfo{person}{Jason Blocklove}, \bibinfo{person}{Siddharth Garg}, \bibinfo{person}{Ramesh Karri}, {and} \bibinfo{person}{Hammond Pearce}.} \bibinfo{year}{2023}\natexlab{}.
\newblock \showarticletitle{Chip-chat: Challenges and opportunities in conversational hardware design}. In \bibinfo{booktitle}{\emph{2023 ACM/IEEE 5th Workshop on Machine Learning for CAD (MLCAD)}}. IEEE, \bibinfo{pages}{1--6}.
\newblock


\bibitem[Chan et~al\mbox{.}(2023)]%
        {chan2023chateval}
\bibfield{author}{\bibinfo{person}{Chi-Min Chan}, \bibinfo{person}{Weize Chen}, \bibinfo{person}{Yusheng Su}, \bibinfo{person}{Jianxuan Yu}, \bibinfo{person}{Wei Xue}, \bibinfo{person}{Shanghang Zhang}, \bibinfo{person}{Jie Fu}, {and} \bibinfo{person}{Zhiyuan Liu}.} \bibinfo{year}{2023}\natexlab{}.
\newblock \showarticletitle{Chateval: Towards better llm-based evaluators through multi-agent debate}.
\newblock \bibinfo{journal}{\emph{arXiv preprint arXiv:2308.07201}} (\bibinfo{year}{2023}).
\newblock


\bibitem[Chen et~al\mbox{.}(2023a)]%
        {chen2023autoagents}
\bibfield{author}{\bibinfo{person}{Guangyao Chen}, \bibinfo{person}{Siwei Dong}, \bibinfo{person}{Yu Shu}, \bibinfo{person}{Ge Zhang}, \bibinfo{person}{Jaward Sesay}, \bibinfo{person}{B{\"o}rje~F Karlsson}, \bibinfo{person}{Jie Fu}, {and} \bibinfo{person}{Yemin Shi}.} \bibinfo{year}{2023}\natexlab{a}.
\newblock \showarticletitle{Autoagents: A framework for automatic agent generation}.
\newblock \bibinfo{journal}{\emph{arXiv preprint arXiv:2309.17288}} (\bibinfo{year}{2023}).
\newblock


\bibitem[Chen et~al\mbox{.}(2023b)]%
        {chen2023agentverse}
\bibfield{author}{\bibinfo{person}{Weize Chen}, \bibinfo{person}{Yusheng Su}, \bibinfo{person}{Jingwei Zuo}, \bibinfo{person}{Cheng Yang}, \bibinfo{person}{Chenfei Yuan}, \bibinfo{person}{Chi-Min Chan}, \bibinfo{person}{Heyang Yu}, \bibinfo{person}{Yaxi Lu}, \bibinfo{person}{Yi-Hsin Hung}, \bibinfo{person}{Chen Qian}, {et~al\mbox{.}}} \bibinfo{year}{2023}\natexlab{b}.
\newblock \showarticletitle{Agentverse: Facilitating multi-agent collaboration and exploring emergent behaviors}. In \bibinfo{booktitle}{\emph{The Twelfth International Conference on Learning Representations}}.
\newblock


\bibitem[Cheng et~al\mbox{.}(2023)]%
        {PromptSapper}
\bibfield{author}{\bibinfo{person}{Yu Cheng}, \bibinfo{person}{Jieshan Chen}, \bibinfo{person}{Qing Huang}, \bibinfo{person}{Zhenchang Xing}, \bibinfo{person}{Xiwei Xu}, {and} \bibinfo{person}{Qinghua Lu}.} \bibinfo{year}{2023}\natexlab{}.
\newblock \showarticletitle{Prompt Sapper: A LLM-Empowered Production Tool for Building AI Chains}.
\newblock  (\bibinfo{date}{dec} \bibinfo{year}{2023}).
\newblock
\showISSN{1049-331X}
\urldef\tempurl%
\url{https://doi.org/10.1145/3638247}
\showDOI{\tempurl}


\bibitem[Dang et~al\mbox{.}(2023)]%
        {dang2023choice}
\bibfield{author}{\bibinfo{person}{Hai Dang}, \bibinfo{person}{Sven Goller}, \bibinfo{person}{Florian Lehmann}, {and} \bibinfo{person}{Daniel Buschek}.} \bibinfo{year}{2023}\natexlab{}.
\newblock \showarticletitle{Choice over control: How users write with large language models using diegetic and non-diegetic prompting}. In \bibinfo{booktitle}{\emph{Proceedings of the 2023 CHI Conference on Human Factors in Computing Systems}}. \bibinfo{pages}{1--17}.
\newblock


\bibitem[Deng et~al\mbox{.}(2022)]%
        {bonbon}
\bibfield{author}{\bibinfo{person}{Jialin Deng}, \bibinfo{person}{Patrick Olivier}, \bibinfo{person}{Josh Andres}, \bibinfo{person}{Kirsten Ellis}, \bibinfo{person}{Ryan Wee}, {and} \bibinfo{person}{Florian Floyd~Mueller}.} \bibinfo{year}{2022}\natexlab{}.
\newblock \showarticletitle{Logic Bonbon: Exploring Food as Computational Artifact}. In \bibinfo{booktitle}{\emph{CHI Conference on Human Factors in Computing Systems}} (New Orleans, LA, USA) \emph{(\bibinfo{series}{CHI '22})}. \bibinfo{publisher}{Association for Computing Machinery}, \bibinfo{address}{New York, NY, USA}, \bibinfo{pages}{1--21}.
\newblock


\bibitem[Dessouky et~al\mbox{.}(2019)]%
        {dessouky2019hardfails}
\bibfield{author}{\bibinfo{person}{Ghada Dessouky}, \bibinfo{person}{David Gens}, \bibinfo{person}{Patrick Haney}, \bibinfo{person}{Garrett Persyn}, \bibinfo{person}{Arun Kanuparthi}, \bibinfo{person}{Hareesh Khattri}, \bibinfo{person}{Jason~M Fung}, \bibinfo{person}{Ahmad-Reza Sadeghi}, {and} \bibinfo{person}{Jeyavijayan Rajendran}.} \bibinfo{year}{2019}\natexlab{}.
\newblock \showarticletitle{$\{$HardFails$\}$: Insights into $\{$Software-Exploitable$\}$ Hardware Bugs}. In \bibinfo{booktitle}{\emph{28th USENIX Security Symposium (USENIX Security 19)}}. \bibinfo{pages}{213--230}.
\newblock


\bibitem[Drew et~al\mbox{.}(2016)]%
        {Toastboard}
\bibfield{author}{\bibinfo{person}{Daniel Drew}, \bibinfo{person}{Julie~L. Newcomb}, \bibinfo{person}{William McGrath}, \bibinfo{person}{Filip Maksimovic}, \bibinfo{person}{David Mellis}, {and} \bibinfo{person}{Bj\"{o}rn Hartmann}.} \bibinfo{year}{2016}\natexlab{}.
\newblock \showarticletitle{The Toastboard: Ubiquitous Instrumentation and Automated Checking of Breadboarded Circuits} \emph{(\bibinfo{series}{UIST '16})}. \bibinfo{publisher}{Association for Computing Machinery}, \bibinfo{address}{New York, NY, USA}, \bibinfo{pages}{677–686}.
\newblock
\showISBNx{9781450341899}
\urldef\tempurl%
\url{https://doi.org/10.1145/2984511.2984566}
\showDOI{\tempurl}


\bibitem[Duan et~al\mbox{.}(2023)]%
        {UIllm}
\bibfield{author}{\bibinfo{person}{Peitong Duan}, \bibinfo{person}{Jeremy Warner}, {and} \bibinfo{person}{Bjoern Hartmann}.} \bibinfo{year}{2023}\natexlab{}.
\newblock \showarticletitle{Towards Generating UI Design Feedback with LLMs} \emph{(\bibinfo{series}{UIST '23 Adjunct})}. \bibinfo{publisher}{Association for Computing Machinery}, \bibinfo{address}{New York, NY, USA}, \bibinfo{numpages}{3}~pages.
\newblock
\showISBNx{9798400700965}
\urldef\tempurl%
\url{https://doi.org/10.1145/3586182.3615810}
\showDOI{\tempurl}


\bibitem[Gmeiner et~al\mbox{.}(2023)]%
        {gmeiner2023exploring}
\bibfield{author}{\bibinfo{person}{Frederic Gmeiner}, \bibinfo{person}{Humphrey Yang}, \bibinfo{person}{Lining Yao}, \bibinfo{person}{Kenneth Holstein}, {and} \bibinfo{person}{Nikolas Martelaro}.} \bibinfo{year}{2023}\natexlab{}.
\newblock \showarticletitle{Exploring challenges and opportunities to support designers in learning to co-create with AI-based manufacturing design tools}. In \bibinfo{booktitle}{\emph{Proceedings of the 2023 CHI Conference on Human Factors in Computing Systems}}. \bibinfo{pages}{1--20}.
\newblock


\bibitem[Gourevitch and Morris(2008)]%
        {gourevitch2008standard}
\bibfield{author}{\bibinfo{person}{Philip Gourevitch} {and} \bibinfo{person}{Errol Morris}.} \bibinfo{year}{2008}\natexlab{}.
\newblock \bibinfo{booktitle}{\emph{Standard operating procedure}}.
\newblock \bibinfo{publisher}{Penguin}.
\newblock


\bibitem[Hauser et~al\mbox{.}(2011)]%
        {1.7}
\bibfield{author}{\bibinfo{person}{Helmut Hauser}, \bibinfo{person}{Auke~J Ijspeert}, \bibinfo{person}{Rudolf~M F{\"u}chslin}, \bibinfo{person}{Rolf Pfeifer}, {and} \bibinfo{person}{Wolfgang Maass}.} \bibinfo{year}{2011}\natexlab{}.
\newblock \showarticletitle{Towards a theoretical foundation for morphological computation with compliant bodies}.
\newblock \bibinfo{journal}{\emph{Biological cybernetics}} \bibinfo{volume}{105}, \bibinfo{number}{5} (\bibinfo{year}{2011}), \bibinfo{pages}{355--370}.
\newblock


\bibitem[Hong et~al\mbox{.}(2023)]%
        {hong2023metagpt}
\bibfield{author}{\bibinfo{person}{Sirui Hong}, \bibinfo{person}{Xiawu Zheng}, \bibinfo{person}{Jonathan Chen}, \bibinfo{person}{Yuheng Cheng}, \bibinfo{person}{Jinlin Wang}, \bibinfo{person}{Ceyao Zhang}, \bibinfo{person}{Zili Wang}, \bibinfo{person}{Steven Ka~Shing Yau}, \bibinfo{person}{Zijuan Lin}, \bibinfo{person}{Liyang Zhou}, {et~al\mbox{.}}} \bibinfo{year}{2023}\natexlab{}.
\newblock \showarticletitle{Metagpt: Meta programming for multi-agent collaborative framework}.
\newblock \bibinfo{journal}{\emph{arXiv preprint arXiv:2308.00352}} (\bibinfo{year}{2023}).
\newblock


\bibitem[Hu et~al\mbox{.}(2024)]%
        {hu2024bliva}
\bibfield{author}{\bibinfo{person}{Wenbo Hu}, \bibinfo{person}{Yifan Xu}, \bibinfo{person}{Yi Li}, \bibinfo{person}{Weiyue Li}, \bibinfo{person}{Zeyuan Chen}, {and} \bibinfo{person}{Zhuowen Tu}.} \bibinfo{year}{2024}\natexlab{}.
\newblock \showarticletitle{Bliva: A simple multimodal llm for better handling of text-rich visual questions}. In \bibinfo{booktitle}{\emph{Proceedings of the AAAI Conference on Artificial Intelligence}}, Vol.~\bibinfo{volume}{38}. \bibinfo{pages}{2256--2264}.
\newblock


\bibitem[Huh et~al\mbox{.}(2023)]%
        {huh2023genassist}
\bibfield{author}{\bibinfo{person}{Mina Huh}, \bibinfo{person}{Yi-Hao Peng}, {and} \bibinfo{person}{Amy Pavel}.} \bibinfo{year}{2023}\natexlab{}.
\newblock \showarticletitle{GenAssist: Making Image Generation Accessible}. In \bibinfo{booktitle}{\emph{Proceedings of the 36th Annual ACM Symposium on User Interface Software and Technology}}. \bibinfo{pages}{1--17}.
\newblock


\bibitem[Ion et~al\mbox{.}(2017a)]%
        {ion_digital_2017}
\bibfield{author}{\bibinfo{person}{Alexandra Ion}, \bibinfo{person}{Ludwig Wall}, \bibinfo{person}{Robert Kovacs}, {and} \bibinfo{person}{Patrick Baudisch}.} \bibinfo{year}{2017}\natexlab{a}.
\newblock \showarticletitle{Digital {Mechanical} {Metamaterials}}. In \bibinfo{booktitle}{\emph{Proceedings of the 2017 {CHI} {Conference} on {Human} {Factors} in {Computing} {Systems}}} \emph{(\bibinfo{series}{{CHI} '17})}. \bibinfo{publisher}{Association for Computing Machinery}, \bibinfo{address}{New York, NY, USA}, \bibinfo{pages}{977--988}.
\newblock
\showISBNx{9781450346559}
\urldef\tempurl%
\url{https://doi.org/10.1145/3025453.3025624}
\showDOI{\tempurl}


\bibitem[Ion et~al\mbox{.}(2017b)]%
        {metamaterial}
\bibfield{author}{\bibinfo{person}{Alexandra Ion}, \bibinfo{person}{Ludwig Wall}, \bibinfo{person}{Robert Kovacs}, {and} \bibinfo{person}{Patrick Baudisch}.} \bibinfo{year}{2017}\natexlab{b}.
\newblock \showarticletitle{Digital mechanical metamaterials}. In \bibinfo{booktitle}{\emph{Proceedings of the 2017 CHI Conference on Human Factors in Computing Systems}} (Denver, Colorado, USA) \emph{(\bibinfo{series}{CHI '17})}. \bibinfo{publisher}{Association for Computing Machinery}, \bibinfo{address}{New York, NY, USA}, \bibinfo{pages}{977--988}.
\newblock


\bibitem[Ion et~al\mbox{.}(2017c)]%
        {DMM}
\bibfield{author}{\bibinfo{person}{Alexandra Ion}, \bibinfo{person}{Ludwig Wall}, \bibinfo{person}{Robert Kovacs}, {and} \bibinfo{person}{Patrick Baudisch}.} \bibinfo{year}{2017}\natexlab{c}.
\newblock \showarticletitle{Digital Mechanical Metamaterials}. In \bibinfo{booktitle}{\emph{Proceedings of the 2017 CHI Conference on Human Factors in Computing Systems}} (Denver, Colorado, USA) \emph{(\bibinfo{series}{CHI '17})}. \bibinfo{publisher}{Association for Computing Machinery}, \bibinfo{address}{New York, NY, USA}, \bibinfo{pages}{977–988}.
\newblock
\showISBNx{9781450346559}
\urldef\tempurl%
\url{https://doi.org/10.1145/3025453.3025624}
\showDOI{\tempurl}


\bibitem[Jain and Kukkal(2020)]%
        {dataCAD}
\bibfield{author}{\bibinfo{person}{Rajeev Jain} {and} \bibinfo{person}{Pankaj Kukkal}.} \bibinfo{year}{2020}\natexlab{}.
\newblock \showarticletitle{Data-driven CAD or Algorithm-Driven CAD: Competitors or Collaborators?} \emph{(\bibinfo{series}{MLCAD '20})}. \bibinfo{publisher}{Association for Computing Machinery}, \bibinfo{address}{New York, NY, USA}, \bibinfo{pages}{69}.
\newblock
\showISBNx{9781450375191}
\urldef\tempurl%
\url{https://doi.org/10.1145/3380446.3430686}
\showDOI{\tempurl}


\bibitem[Jennings et~al\mbox{.}(1998)]%
        {jennings1998roadmap}
\bibfield{author}{\bibinfo{person}{Nicholas~R Jennings}, \bibinfo{person}{Katia Sycara}, {and} \bibinfo{person}{Michael Wooldridge}.} \bibinfo{year}{1998}\natexlab{}.
\newblock \showarticletitle{A roadmap of agent research and development}.
\newblock \bibinfo{journal}{\emph{Autonomous agents and multi-agent systems}}  \bibinfo{volume}{1} (\bibinfo{year}{1998}), \bibinfo{pages}{7--38}.
\newblock


\bibitem[Jiang et~al\mbox{.}(2023)]%
        {jiang2023graphologue}
\bibfield{author}{\bibinfo{person}{Peiling Jiang}, \bibinfo{person}{Jude Rayan}, \bibinfo{person}{Steven~P Dow}, {and} \bibinfo{person}{Haijun Xia}.} \bibinfo{year}{2023}\natexlab{}.
\newblock \showarticletitle{Graphologue: Exploring large language model responses with interactive diagrams}. In \bibinfo{booktitle}{\emph{Proceedings of the 36th Annual ACM Symposium on User Interface Software and Technology}}. \bibinfo{pages}{1--20}.
\newblock


\bibitem[Jo et~al\mbox{.}(2023)]%
        {jo2023understanding}
\bibfield{author}{\bibinfo{person}{Eunkyung Jo}, \bibinfo{person}{Daniel~A Epstein}, \bibinfo{person}{Hyunhoon Jung}, {and} \bibinfo{person}{Young-Ho Kim}.} \bibinfo{year}{2023}\natexlab{}.
\newblock \showarticletitle{Understanding the benefits and challenges of deploying conversational AI leveraging large language models for public health intervention}. In \bibinfo{booktitle}{\emph{Proceedings of the 2023 CHI Conference on Human Factors in Computing Systems}}. \bibinfo{pages}{1--16}.
\newblock


\bibitem[Jun and Nichol(2023)]%
        {jun2023shap}
\bibfield{author}{\bibinfo{person}{Heewoo Jun} {and} \bibinfo{person}{Alex Nichol}.} \bibinfo{year}{2023}\natexlab{}.
\newblock \showarticletitle{Shap-e: Generating conditional 3d implicit functions}.
\newblock \bibinfo{journal}{\emph{arXiv preprint arXiv:2305.02463}} (\bibinfo{year}{2023}).
\newblock


\bibitem[Ko et~al\mbox{.}(2023)]%
        {t2image}
\bibfield{author}{\bibinfo{person}{Hyung-Kwon Ko}, \bibinfo{person}{Gwanmo Park}, \bibinfo{person}{Hyeon Jeon}, \bibinfo{person}{Jaemin Jo}, \bibinfo{person}{Juho Kim}, {and} \bibinfo{person}{Jinwook Seo}.} \bibinfo{year}{2023}\natexlab{}.
\newblock \showarticletitle{Large-scale Text-to-Image Generation Models for Visual Artists’ Creative Works} \emph{(\bibinfo{series}{IUI '23})}. \bibinfo{publisher}{Association for Computing Machinery}, \bibinfo{address}{New York, NY, USA}, \bibinfo{numpages}{15}~pages.
\newblock
\showISBNx{9798400701061}
\urldef\tempurl%
\url{https://doi.org/10.1145/3581641.3584078}
\showDOI{\tempurl}


\bibitem[Li et~al\mbox{.}(2024)]%
        {li2024camel}
\bibfield{author}{\bibinfo{person}{Guohao Li}, \bibinfo{person}{Hasan Hammoud}, \bibinfo{person}{Hani Itani}, \bibinfo{person}{Dmitrii Khizbullin}, {and} \bibinfo{person}{Bernard Ghanem}.} \bibinfo{year}{2024}\natexlab{}.
\newblock \showarticletitle{Camel: Communicative agents for" mind" exploration of large language model society}.
\newblock \bibinfo{journal}{\emph{Advances in Neural Information Processing Systems}}  \bibinfo{volume}{36} (\bibinfo{year}{2024}).
\newblock


\bibitem[Li(2024)]%
        {li2024study}
\bibfield{author}{\bibinfo{person}{Weiyi Li}.} \bibinfo{year}{2024}\natexlab{}.
\newblock \showarticletitle{A Study on Factors Influencing Designers’ Behavioral Intention in Using AI-Generated Content for Assisted Design: Perceived Anxiety, Perceived Risk, and UTAUT}.
\newblock \bibinfo{journal}{\emph{International Journal of Human--Computer Interaction}} (\bibinfo{year}{2024}), \bibinfo{pages}{1--14}.
\newblock


\bibitem[Li et~al\mbox{.}(2023)]%
        {li2023metaagents}
\bibfield{author}{\bibinfo{person}{Yuan Li}, \bibinfo{person}{Yixuan Zhang}, {and} \bibinfo{person}{Lichao Sun}.} \bibinfo{year}{2023}\natexlab{}.
\newblock \showarticletitle{Metaagents: Simulating interactions of human behaviors for llm-based task-oriented coordination via collaborative generative agents}.
\newblock \bibinfo{journal}{\emph{arXiv preprint arXiv:2310.06500}} (\bibinfo{year}{2023}).
\newblock


\bibitem[Liu et~al\mbox{.}(2023a)]%
        {liu2023wants}
\bibfield{author}{\bibinfo{person}{Michael~Xieyang Liu}, \bibinfo{person}{Advait Sarkar}, \bibinfo{person}{Carina Negreanu}, \bibinfo{person}{Benjamin Zorn}, \bibinfo{person}{Jack Williams}, \bibinfo{person}{Neil Toronto}, {and} \bibinfo{person}{Andrew~D Gordon}.} \bibinfo{year}{2023}\natexlab{a}.
\newblock \showarticletitle{“What it wants me to say”: Bridging the abstraction gap between end-user programmers and code-generating large language models}. In \bibinfo{booktitle}{\emph{Proceedings of the 2023 CHI Conference on Human Factors in Computing Systems}}. \bibinfo{pages}{1--31}.
\newblock


\bibitem[Liu et~al\mbox{.}(2023b)]%
        {liu20233dall}
\bibfield{author}{\bibinfo{person}{Vivian Liu}, \bibinfo{person}{Jo Vermeulen}, \bibinfo{person}{George Fitzmaurice}, {and} \bibinfo{person}{Justin Matejka}.} \bibinfo{year}{2023}\natexlab{b}.
\newblock \showarticletitle{3DALL-E: Integrating text-to-image AI in 3D design workflows}. In \bibinfo{booktitle}{\emph{Proceedings of the 2023 ACM designing interactive systems conference}}. \bibinfo{pages}{1955--1977}.
\newblock


\bibitem[Liu et~al\mbox{.}(2024)]%
        {liu2024towards}
\bibfield{author}{\bibinfo{person}{Yuchi Liu}, \bibinfo{person}{Jaskirat Singh}, \bibinfo{person}{Gaowen Liu}, \bibinfo{person}{Ali Payani}, {and} \bibinfo{person}{Liang Zheng}.} \bibinfo{year}{2024}\natexlab{}.
\newblock \showarticletitle{Towards Hierarchical Multi-Agent Workflows for Zero-Shot Prompt Optimization}.
\newblock \bibinfo{journal}{\emph{arXiv preprint arXiv:2405.20252}} (\bibinfo{year}{2024}).
\newblock


\bibitem[Lu et~al\mbox{.}(2023b)]%
        {lu_fluidic_2023}
\bibfield{author}{\bibinfo{person}{Qiuyu Lu}, \bibinfo{person}{Haiqing Xu}, \bibinfo{person}{Yijie Guo}, \bibinfo{person}{Joey~Yu Wang}, {and} \bibinfo{person}{Lining Yao}.} \bibinfo{year}{2023}\natexlab{b}.
\newblock \showarticletitle{Fluidic {Computation} {Kit}: {Towards} {Electronic}-free {Shape}-changing {Interfaces}}. In \bibinfo{booktitle}{\emph{Proceedings of the 2023 {CHI} {Conference} on {Human} {Factors} in {Computing} {Systems}}} \emph{(\bibinfo{series}{{CHI} '23})}. \bibinfo{publisher}{Association for Computing Machinery}, \bibinfo{address}{New York, NY, USA}, \bibinfo{pages}{1--21}.
\newblock
\showISBNx{9781450394215}
\urldef\tempurl%
\url{https://doi.org/10.1145/3544548.3580783}
\showDOI{\tempurl}


\bibitem[Lu et~al\mbox{.}(2023c)]%
        {Sustainflatable}
\bibfield{author}{\bibinfo{person}{Qiuyu Lu}, \bibinfo{person}{Tianyu Yu}, \bibinfo{person}{Semina Yi}, \bibinfo{person}{Yuran Ding}, \bibinfo{person}{Haipeng Mi}, {and} \bibinfo{person}{Lining Yao}.} \bibinfo{year}{2023}\natexlab{c}.
\newblock \showarticletitle{Sustainflatable: Harvesting, Storing and Utilizing Ambient Energy for Pneumatic Morphing Interfaces}. In \bibinfo{booktitle}{\emph{Proceedings of the 36th Annual ACM Symposium on User Interface Software and Technology}} (<conf-loc>, <city>San Francisco</city>, <state>CA</state>, <country>USA</country>, </conf-loc>) \emph{(\bibinfo{series}{UIST '23})}. \bibinfo{publisher}{Association for Computing Machinery}, \bibinfo{address}{New York, NY, USA}, Article \bibinfo{articleno}{32}, \bibinfo{numpages}{20}~pages.
\newblock
\showISBNx{9798400701320}
\urldef\tempurl%
\url{https://doi.org/10.1145/3586183.3606721}
\showDOI{\tempurl}


\bibitem[Lu et~al\mbox{.}(2023a)]%
        {lu2023readingquizmaker}
\bibfield{author}{\bibinfo{person}{Xinyi Lu}, \bibinfo{person}{Simin Fan}, \bibinfo{person}{Jessica Houghton}, \bibinfo{person}{Lu Wang}, {and} \bibinfo{person}{Xu Wang}.} \bibinfo{year}{2023}\natexlab{a}.
\newblock \showarticletitle{Readingquizmaker: A human-nlp collaborative system that supports instructors to design high-quality reading quiz questions}. In \bibinfo{booktitle}{\emph{Proceedings of the 2023 CHI Conference on Human Factors in Computing Systems}}. \bibinfo{pages}{1--18}.
\newblock


\bibitem[Miller(2019)]%
        {miller2019explanation}
\bibfield{author}{\bibinfo{person}{Tim Miller}.} \bibinfo{year}{2019}\natexlab{}.
\newblock \showarticletitle{Explanation in artificial intelligence: Insights from the social sciences}.
\newblock \bibinfo{journal}{\emph{Artificial intelligence}}  \bibinfo{volume}{267} (\bibinfo{year}{2019}), \bibinfo{pages}{1--38}.
\newblock


\bibitem[Mor et~al\mbox{.}(2020)]%
        {venous}
\bibfield{author}{\bibinfo{person}{Hila Mor}, \bibinfo{person}{Tianyu Yu}, \bibinfo{person}{Ken Nakagaki}, \bibinfo{person}{Benjamin~Harvey Miller}, \bibinfo{person}{Yichen Jia}, {and} \bibinfo{person}{Hiroshi Ishii}.} \bibinfo{year}{2020}\natexlab{}.
\newblock \showarticletitle{Venous Materials: Towards Interactive Fluidic Mechanisms}. In \bibinfo{booktitle}{\emph{Proceedings of the 2020 CHI Conference on Human Factors in Computing Systems}} (Honolulu, HI, USA) \emph{(\bibinfo{series}{CHI '20})}. \bibinfo{publisher}{Association for Computing Machinery}, \bibinfo{address}{New York, NY, USA}, \bibinfo{pages}{1–14}.
\newblock
\showISBNx{9781450367080}
\urldef\tempurl%
\url{https://doi.org/10.1145/3313831.3376129}
\showDOI{\tempurl}


\bibitem[Mosadegh et~al\mbox{.}(2011)]%
        {Next-generation}
\bibfield{author}{\bibinfo{person}{Bobak Mosadegh}, \bibinfo{person}{Tommaso Bersano-Begey}, \bibinfo{person}{Joong~Yull Park}, \bibinfo{person}{Mark~A Burns}, {and} \bibinfo{person}{Shuichi Takayama}.} \bibinfo{year}{2011}\natexlab{}.
\newblock \showarticletitle{Next-generation integrated microfluidic circuits}.
\newblock \bibinfo{journal}{\emph{Lab on a Chip}} \bibinfo{volume}{11}, \bibinfo{number}{17} (\bibinfo{year}{2011}), \bibinfo{pages}{2813--2818}.
\newblock


\bibitem[Nakagaki et~al\mbox{.}(2015)]%
        {lineform}
\bibfield{author}{\bibinfo{person}{Ken Nakagaki}, \bibinfo{person}{Sean Follmer}, {and} \bibinfo{person}{Hiroshi Ishii}.} \bibinfo{year}{2015}\natexlab{}.
\newblock \showarticletitle{LineFORM: Actuated Curve Interfaces for Display, Interaction, and Constraint}. In \bibinfo{booktitle}{\emph{Proceedings of the 28th Annual ACM Symposium on User Interface Software \& Technology}} (Charlotte, NC, USA) \emph{(\bibinfo{series}{UIST '15})}. \bibinfo{publisher}{Association for Computing Machinery}, \bibinfo{address}{New York, NY, USA}, \bibinfo{pages}{333–339}.
\newblock
\showISBNx{9781450337793}
\urldef\tempurl%
\url{https://doi.org/10.1145/2807442.2807452}
\showDOI{\tempurl}


\bibitem[Nakamaru et~al\mbox{.}(2017)]%
        {FoamSense}
\bibfield{author}{\bibinfo{person}{Satoshi Nakamaru}, \bibinfo{person}{Ryosuke Nakayama}, \bibinfo{person}{Ryuma Niiyama}, {and} \bibinfo{person}{Yasuaki Kakehi}.} \bibinfo{year}{2017}\natexlab{}.
\newblock \showarticletitle{FoamSense: Design of Three Dimensional Soft Sensors with Porous Materials}. In \bibinfo{booktitle}{\emph{Proceedings of the 30th Annual ACM Symposium on User Interface Software and Technology}} (Qu\'{e}bec City, QC, Canada) \emph{(\bibinfo{series}{UIST '17})}. \bibinfo{publisher}{Association for Computing Machinery}, \bibinfo{address}{New York, NY, USA}, \bibinfo{pages}{437–447}.
\newblock
\showISBNx{9781450349819}
\urldef\tempurl%
\url{https://doi.org/10.1145/3126594.3126666}
\showDOI{\tempurl}


\bibitem[Nam et~al\mbox{.}(2024)]%
        {nam2024using}
\bibfield{author}{\bibinfo{person}{Daye Nam}, \bibinfo{person}{Andrew Macvean}, \bibinfo{person}{Vincent Hellendoorn}, \bibinfo{person}{Bogdan Vasilescu}, {and} \bibinfo{person}{Brad Myers}.} \bibinfo{year}{2024}\natexlab{}.
\newblock \showarticletitle{Using an llm to help with code understanding}. In \bibinfo{booktitle}{\emph{Proceedings of the IEEE/ACM 46th International Conference on Software Engineering}}. \bibinfo{pages}{1--13}.
\newblock


\bibitem[Olberding et~al\mbox{.}(2014)]%
        {PrintScreen}
\bibfield{author}{\bibinfo{person}{Simon Olberding}, \bibinfo{person}{Michael Wessely}, {and} \bibinfo{person}{J\"{u}rgen Steimle}.} \bibinfo{year}{2014}\natexlab{}.
\newblock \showarticletitle{PrintScreen: fabricating highly customizable thin-film touch-displays}. In \bibinfo{booktitle}{\emph{Proceedings of the 27th Annual ACM Symposium on User Interface Software and Technology}} (Honolulu, Hawaii, USA) \emph{(\bibinfo{series}{UIST '14})}. \bibinfo{publisher}{Association for Computing Machinery}, \bibinfo{address}{New York, NY, USA}, \bibinfo{pages}{281–290}.
\newblock
\showISBNx{9781450330695}
\urldef\tempurl%
\url{https://doi.org/10.1145/2642918.2647413}
\showDOI{\tempurl}


\bibitem[Ou et~al\mbox{.}(2016)]%
        {ou_aeromorph_2016}
\bibfield{author}{\bibinfo{person}{Jifei Ou}, \bibinfo{person}{Mélina Skouras}, \bibinfo{person}{Nikolaos Vlavianos}, \bibinfo{person}{Felix Heibeck}, \bibinfo{person}{Chin-Yi Cheng}, \bibinfo{person}{Jannik Peters}, {and} \bibinfo{person}{Hiroshi Ishii}.} \bibinfo{year}{2016}\natexlab{}.
\newblock \showarticletitle{{aeroMorph} - {Heat}-sealing {Inflatable} {Shape}-change {Materials} for {Interaction} {Design}}. In \bibinfo{booktitle}{\emph{Proceedings of the 29th {Annual} {Symposium} on {User} {Interface} {Software} and {Technology}}} \emph{(\bibinfo{series}{{UIST} '16})}. \bibinfo{publisher}{Association for Computing Machinery}, \bibinfo{address}{New York, NY, USA}, \bibinfo{pages}{121--132}.
\newblock
\showISBNx{9781450341899}
\urldef\tempurl%
\url{https://doi.org/10.1145/2984511.2984520}
\showDOI{\tempurl}


\bibitem[Park et~al\mbox{.}(2023)]%
        {park2023generative}
\bibfield{author}{\bibinfo{person}{Joon~Sung Park}, \bibinfo{person}{Joseph O'Brien}, \bibinfo{person}{Carrie~Jun Cai}, \bibinfo{person}{Meredith~Ringel Morris}, \bibinfo{person}{Percy Liang}, {and} \bibinfo{person}{Michael~S Bernstein}.} \bibinfo{year}{2023}\natexlab{}.
\newblock \showarticletitle{Generative agents: Interactive simulacra of human behavior}. In \bibinfo{booktitle}{\emph{Proceedings of the 36th Annual ACM Symposium on User Interface Software and Technology}}. \bibinfo{pages}{1--22}.
\newblock


\bibitem[Peng et~al\mbox{.}(2018)]%
        {RoMA}
\bibfield{author}{\bibinfo{person}{Huaishu Peng}, \bibinfo{person}{Jimmy Briggs}, \bibinfo{person}{Cheng-Yao Wang}, \bibinfo{person}{Kevin Guo}, \bibinfo{person}{Joseph Kider}, \bibinfo{person}{Stefanie Mueller}, \bibinfo{person}{Patrick Baudisch}, {and} \bibinfo{person}{Fran\c{c}ois Guimbreti\`{e}re}.} \bibinfo{year}{2018}\natexlab{}.
\newblock \showarticletitle{RoMA: Interactive Fabrication with Augmented Reality and a Robotic 3D Printer} \emph{(\bibinfo{series}{CHI '18})}. \bibinfo{publisher}{Association for Computing Machinery}, \bibinfo{address}{New York, NY, USA}, \bibinfo{pages}{1–12}.
\newblock
\showISBNx{9781450356206}
\urldef\tempurl%
\url{https://doi.org/10.1145/3173574.3174153}
\showDOI{\tempurl}


\bibitem[Petridis et~al\mbox{.}(2023)]%
        {petridis2023anglekindling}
\bibfield{author}{\bibinfo{person}{Savvas Petridis}, \bibinfo{person}{Nicholas Diakopoulos}, \bibinfo{person}{Kevin Crowston}, \bibinfo{person}{Mark Hansen}, \bibinfo{person}{Keren Henderson}, \bibinfo{person}{Stan Jastrzebski}, \bibinfo{person}{Jeffrey~V Nickerson}, {and} \bibinfo{person}{Lydia~B Chilton}.} \bibinfo{year}{2023}\natexlab{}.
\newblock \showarticletitle{Anglekindling: Supporting journalistic angle ideation with large language models}. In \bibinfo{booktitle}{\emph{Proceedings of the 2023 CHI Conference on Human Factors in Computing Systems}}. \bibinfo{pages}{1--16}.
\newblock


\bibitem[Preston et~al\mbox{.}(2019a)]%
        {preston_soft_2019}
\bibfield{author}{\bibinfo{person}{Daniel~J. Preston}, \bibinfo{person}{Haihui~Joy Jiang}, \bibinfo{person}{Vanessa Sanchez}, \bibinfo{person}{Philipp Rothemund}, \bibinfo{person}{Jeff Rawson}, \bibinfo{person}{Markus~P. Nemitz}, \bibinfo{person}{Won-Kyu Lee}, \bibinfo{person}{Zhigang Suo}, \bibinfo{person}{Conor~J. Walsh}, {and} \bibinfo{person}{George~M. Whitesides}.} \bibinfo{year}{2019}\natexlab{a}.
\newblock \showarticletitle{A soft ring oscillator}.
\newblock \bibinfo{journal}{\emph{Science Robotics}} \bibinfo{volume}{4}, \bibinfo{number}{31} (\bibinfo{date}{June} \bibinfo{year}{2019}), \bibinfo{pages}{eaaw5496}.
\newblock
\urldef\tempurl%
\url{https://doi.org/10.1126/scirobotics.aaw5496}
\showDOI{\tempurl}


\bibitem[Preston et~al\mbox{.}(2019b)]%
        {preston_digital_2019}
\bibfield{author}{\bibinfo{person}{Daniel~J. Preston}, \bibinfo{person}{Philipp Rothemund}, \bibinfo{person}{Haihui~Joy Jiang}, \bibinfo{person}{Markus~P. Nemitz}, \bibinfo{person}{Jeff Rawson}, \bibinfo{person}{Zhigang Suo}, {and} \bibinfo{person}{George~M. Whitesides}.} \bibinfo{year}{2019}\natexlab{b}.
\newblock \showarticletitle{Digital logic for soft devices}.
\newblock \bibinfo{journal}{\emph{Proceedings of the National Academy of Sciences}} \bibinfo{volume}{116}, \bibinfo{number}{16} (\bibinfo{date}{April} \bibinfo{year}{2019}), \bibinfo{pages}{7750--7759}.
\newblock
\urldef\tempurl%
\url{https://doi.org/10.1073/pnas.1820672116}
\showDOI{\tempurl}


\bibitem[Qian et~al\mbox{.}(2023)]%
        {qian2023communicative}
\bibfield{author}{\bibinfo{person}{Chen Qian}, \bibinfo{person}{Xin Cong}, \bibinfo{person}{Cheng Yang}, \bibinfo{person}{Weize Chen}, \bibinfo{person}{Yusheng Su}, \bibinfo{person}{Juyuan Xu}, \bibinfo{person}{Zhiyuan Liu}, {and} \bibinfo{person}{Maosong Sun}.} \bibinfo{year}{2023}\natexlab{}.
\newblock \showarticletitle{Communicative agents for software development}.
\newblock \bibinfo{journal}{\emph{arXiv preprint arXiv:2307.07924}} (\bibinfo{year}{2023}).
\newblock


\bibitem[Rajappan et~al\mbox{.}(2021)]%
        {1.5}
\bibfield{author}{\bibinfo{person}{Anoop Rajappan}, \bibinfo{person}{Barclay Jumet}, {and} \bibinfo{person}{Daniel~J Preston}.} \bibinfo{year}{2021}\natexlab{}.
\newblock \showarticletitle{Pneumatic soft robots take a step toward autonomy}.
\newblock \bibinfo{journal}{\emph{Science Robotics}} \bibinfo{volume}{6}, \bibinfo{number}{51} (\bibinfo{year}{2021}), \bibinfo{pages}{eabg6994}.
\newblock


\bibitem[Rajappan et~al\mbox{.}(2022)]%
        {rajappan_logic-enabled_2022}
\bibfield{author}{\bibinfo{person}{Anoop Rajappan}, \bibinfo{person}{Barclay Jumet}, \bibinfo{person}{Rachel~A. Shveda}, \bibinfo{person}{Colter~J. Decker}, \bibinfo{person}{Zhen Liu}, \bibinfo{person}{Te~Faye Yap}, \bibinfo{person}{Vanessa Sanchez}, {and} \bibinfo{person}{Daniel~J. Preston}.} \bibinfo{year}{2022}\natexlab{}.
\newblock \showarticletitle{Logic-enabled textiles}.
\newblock \bibinfo{journal}{\emph{Proceedings of the National Academy of Sciences}} \bibinfo{volume}{119}, \bibinfo{number}{35} (\bibinfo{date}{Aug.} \bibinfo{year}{2022}), \bibinfo{pages}{e2202118119}.
\newblock
\urldef\tempurl%
\url{https://doi.org/10.1073/pnas.2202118119}
\showDOI{\tempurl}


\bibitem[Rhee and Burns(2009)]%
        {1st-pneumatic-logic-circuits}
\bibfield{author}{\bibinfo{person}{Minsoung Rhee} {and} \bibinfo{person}{Mark~A Burns}.} \bibinfo{year}{2009}\natexlab{}.
\newblock \showarticletitle{Microfluidic pneumatic logic circuits and digital pneumatic microprocessors for integrated microfluidic systems}.
\newblock \bibinfo{journal}{\emph{Lab on a Chip}} \bibinfo{volume}{9}, \bibinfo{number}{21} (\bibinfo{year}{2009}), \bibinfo{pages}{3131--3143}.
\newblock


\bibitem[Savage et~al\mbox{.}(2022)]%
        {savage_airlogic_2022}
\bibfield{author}{\bibinfo{person}{Valkyrie Savage}, \bibinfo{person}{Carlos Tejada}, \bibinfo{person}{Mengyu Zhong}, \bibinfo{person}{Raf Ramakers}, \bibinfo{person}{Daniel Ashbrook}, {and} \bibinfo{person}{Hyunyoung Kim}.} \bibinfo{year}{2022}\natexlab{}.
\newblock \showarticletitle{{AirLogic}: {Embedding} {Pneumatic} {Computation} and {I}/{O} in {3D} {Models} to {Fabricate} {Electronics}-{Free} {Interactive} {Objects}}. In \bibinfo{booktitle}{\emph{Proceedings of the 35th {Annual} {ACM} {Symposium} on {User} {Interface} {Software} and {Technology}}} \emph{(\bibinfo{series}{{UIST} '22})}. \bibinfo{publisher}{Association for Computing Machinery}, \bibinfo{address}{New York, NY, USA}, \bibinfo{pages}{1--12}.
\newblock
\showISBNx{9781450393201}
\urldef\tempurl%
\url{https://doi.org/10.1145/3526113.3545642}
\showDOI{\tempurl}


\bibitem[Shi et~al\mbox{.}(2023)]%
        {shi2023mvdream}
\bibfield{author}{\bibinfo{person}{Yichun Shi}, \bibinfo{person}{Peng Wang}, \bibinfo{person}{Jianglong Ye}, \bibinfo{person}{Mai Long}, \bibinfo{person}{Kejie Li}, {and} \bibinfo{person}{Xiao Yang}.} \bibinfo{year}{2023}\natexlab{}.
\newblock \showarticletitle{Mvdream: Multi-view diffusion for 3d generation}.
\newblock \bibinfo{journal}{\emph{arXiv preprint arXiv:2308.16512}} (\bibinfo{year}{2023}).
\newblock


\bibitem[Shveda et~al\mbox{.}(2022)]%
        {shveda_wearable_2022}
\bibfield{author}{\bibinfo{person}{Rachel~A. Shveda}, \bibinfo{person}{Anoop Rajappan}, \bibinfo{person}{Te~Faye Yap}, \bibinfo{person}{Zhen Liu}, \bibinfo{person}{Marquise~D. Bell}, \bibinfo{person}{Barclay Jumet}, \bibinfo{person}{Vanessa Sanchez}, {and} \bibinfo{person}{Daniel~J. Preston}.} \bibinfo{year}{2022}\natexlab{}.
\newblock \showarticletitle{A wearable textile-based pneumatic energy harvesting system for assistive robotics}.
\newblock \bibinfo{journal}{\emph{Science Advances}} \bibinfo{volume}{8}, \bibinfo{number}{34} (\bibinfo{date}{Aug.} \bibinfo{year}{2022}), \bibinfo{pages}{eabo2418}.
\newblock
\urldef\tempurl%
\url{https://doi.org/10.1126/sciadv.abo2418}
\showDOI{\tempurl}


\bibitem[Suh et~al\mbox{.}(2023)]%
        {suh2023sensecape}
\bibfield{author}{\bibinfo{person}{Sangho Suh}, \bibinfo{person}{Bryan Min}, \bibinfo{person}{Srishti Palani}, {and} \bibinfo{person}{Haijun Xia}.} \bibinfo{year}{2023}\natexlab{}.
\newblock \showarticletitle{Sensecape: Enabling multilevel exploration and sensemaking with large language models}. In \bibinfo{booktitle}{\emph{Proceedings of the 36th Annual ACM Symposium on User Interface Software and Technology}}. \bibinfo{pages}{1--18}.
\newblock


\bibitem[Sun et~al\mbox{.}(2022)]%
        {XBridges}
\bibfield{author}{\bibinfo{person}{Lingyun Sun}, \bibinfo{person}{Jiaji Li}, \bibinfo{person}{Junzhe Ji}, \bibinfo{person}{Deying Pan}, \bibinfo{person}{Mingming Li}, \bibinfo{person}{Kuangqi Zhu}, \bibinfo{person}{Yitao Fan}, \bibinfo{person}{Yue Yang}, \bibinfo{person}{Ye Tao}, {and} \bibinfo{person}{Guanyun Wang}.} \bibinfo{year}{2022}\natexlab{}.
\newblock \showarticletitle{X-Bridges: Designing Tunable Bridges to Enrich 3D Printed Objects' Deformation and Stiffness} \emph{(\bibinfo{series}{UIST '22})}. \bibinfo{publisher}{Association for Computing Machinery}, \bibinfo{address}{New York, NY, USA}, \bibinfo{numpages}{12}~pages.
\newblock
\showISBNx{9781450393201}
\urldef\tempurl%
\url{https://doi.org/10.1145/3526113.3545710}
\showDOI{\tempurl}


\bibitem[Tao et~al\mbox{.}(2023)]%
        {4Doodle}
\bibfield{author}{\bibinfo{person}{Ye Tao}, \bibinfo{person}{Shuhong Wang}, \bibinfo{person}{Junzhe Ji}, \bibinfo{person}{Linlin Cai}, \bibinfo{person}{Hongmei Xia}, \bibinfo{person}{Zhiqi Wang}, \bibinfo{person}{Jinghai He}, \bibinfo{person}{Yitao Fan}, \bibinfo{person}{Shengzhang Pan}, \bibinfo{person}{Jinghua Xu}, \bibinfo{person}{Cheng Yang}, \bibinfo{person}{Lingyun Sun}, {and} \bibinfo{person}{Guanyun Wang}.} \bibinfo{year}{2023}\natexlab{}.
\newblock \showarticletitle{4Doodle: 4D Printing Artifacts Without 3D Printers} \emph{(\bibinfo{series}{CHI '23})}. \bibinfo{publisher}{Association for Computing Machinery}, \bibinfo{address}{New York, NY, USA}, Article \bibinfo{articleno}{731}, \bibinfo{numpages}{16}~pages.
\newblock
\showISBNx{9781450394215}
\urldef\tempurl%
\url{https://doi.org/10.1145/3544548.3581321}
\showDOI{\tempurl}


\bibitem[Teng et~al\mbox{.}(2018)]%
        {teng_pupop_2018}
\bibfield{author}{\bibinfo{person}{Shan-Yuan Teng}, \bibinfo{person}{Tzu-Sheng Kuo}, \bibinfo{person}{Chi Wang}, \bibinfo{person}{Chi-huan Chiang}, \bibinfo{person}{Da-Yuan Huang}, \bibinfo{person}{Liwei Chan}, {and} \bibinfo{person}{Bing-Yu Chen}.} \bibinfo{year}{2018}\natexlab{}.
\newblock \showarticletitle{{PuPoP}: {Pop}-up {Prop} on {Palm} for {Virtual} {Reality}}. In \bibinfo{booktitle}{\emph{Proceedings of the 31st {Annual} {ACM} {Symposium} on {User} {Interface} {Software} and {Technology}}} \emph{(\bibinfo{series}{{UIST} '18})}. \bibinfo{publisher}{Association for Computing Machinery}, \bibinfo{address}{New York, NY, USA}, \bibinfo{pages}{5--17}.
\newblock
\showISBNx{9781450359481}
\urldef\tempurl%
\url{https://doi.org/10.1145/3242587.3242628}
\showDOI{\tempurl}


\bibitem[Tseng et~al\mbox{.}(2024)]%
        {tseng2024keyframer}
\bibfield{author}{\bibinfo{person}{Tiffany Tseng}, \bibinfo{person}{Ruijia Cheng}, {and} \bibinfo{person}{Jeffrey Nichols}.} \bibinfo{year}{2024}\natexlab{}.
\newblock \showarticletitle{Keyframer: Empowering Animation Design using Large Language Models}.
\newblock \bibinfo{journal}{\emph{arXiv preprint arXiv:2402.06071}} (\bibinfo{year}{2024}).
\newblock


\bibitem[Valencia et~al\mbox{.}(2023)]%
        {valencia2023less}
\bibfield{author}{\bibinfo{person}{Stephanie Valencia}, \bibinfo{person}{Richard Cave}, \bibinfo{person}{Krystal Kallarackal}, \bibinfo{person}{Katie Seaver}, \bibinfo{person}{Michael Terry}, {and} \bibinfo{person}{Shaun~K Kane}.} \bibinfo{year}{2023}\natexlab{}.
\newblock \showarticletitle{“The less I type, the better”: How AI Language Models can Enhance or Impede Communication for AAC Users}. In \bibinfo{booktitle}{\emph{Proceedings of the 2023 CHI Conference on Human Factors in Computing Systems}}. \bibinfo{pages}{1--14}.
\newblock


\bibitem[Vaswani et~al\mbox{.}(2017)]%
        {vaswani2017attention}
\bibfield{author}{\bibinfo{person}{Ashish Vaswani}, \bibinfo{person}{Noam Shazeer}, \bibinfo{person}{Niki Parmar}, \bibinfo{person}{Jakob Uszkoreit}, \bibinfo{person}{Llion Jones}, \bibinfo{person}{Aidan~N Gomez}, \bibinfo{person}{{\L}ukasz Kaiser}, {and} \bibinfo{person}{Illia Polosukhin}.} \bibinfo{year}{2017}\natexlab{}.
\newblock \showarticletitle{Attention is all you need}.
\newblock \bibinfo{journal}{\emph{Advances in neural information processing systems}}  \bibinfo{volume}{30} (\bibinfo{year}{2017}).
\newblock


\bibitem[Vere and Bickmore(1990)]%
        {vere1990basic}
\bibfield{author}{\bibinfo{person}{Steven Vere} {and} \bibinfo{person}{Timothy Bickmore}.} \bibinfo{year}{1990}\natexlab{}.
\newblock \showarticletitle{A basic agent}.
\newblock \bibinfo{journal}{\emph{Computational intelligence}} \bibinfo{volume}{6}, \bibinfo{number}{1} (\bibinfo{year}{1990}), \bibinfo{pages}{41--60}.
\newblock


\bibitem[Villar et~al\mbox{.}(2018)]%
        {Zanzibar}
\bibfield{author}{\bibinfo{person}{Nicolas Villar}, \bibinfo{person}{Daniel Cletheroe}, \bibinfo{person}{Greg Saul}, \bibinfo{person}{Christian Holz}, \bibinfo{person}{Tim Regan}, \bibinfo{person}{Oscar Salandin}, \bibinfo{person}{Misha Sra}, \bibinfo{person}{Hui-Shyong Yeo}, \bibinfo{person}{William Field}, {and} \bibinfo{person}{Haiyan Zhang}.} \bibinfo{year}{2018}\natexlab{}.
\newblock \showarticletitle{Project Zanzibar: A Portable and Flexible Tangible Interaction Platform} \emph{(\bibinfo{series}{CHI '18})}. \bibinfo{publisher}{Association for Computing Machinery}, \bibinfo{address}{New York, NY, USA}.
\newblock
\showISBNx{9781450356206}
\urldef\tempurl%
\url{https://doi.org/10.1145/3173574.3174089}
\showDOI{\tempurl}


\bibitem[Wang et~al\mbox{.}(2023a)]%
        {wang2023survey}
\bibfield{author}{\bibinfo{person}{Lei Wang}, \bibinfo{person}{Chen Ma}, \bibinfo{person}{Xueyang Feng}, \bibinfo{person}{Zeyu Zhang}, \bibinfo{person}{Hao Yang}, \bibinfo{person}{Jingsen Zhang}, \bibinfo{person}{Zhiyuan Chen}, \bibinfo{person}{Jiakai Tang}, \bibinfo{person}{Xu Chen}, \bibinfo{person}{Yankai Lin}, {et~al\mbox{.}}} \bibinfo{year}{2023}\natexlab{a}.
\newblock \showarticletitle{A survey on large language model based autonomous agents}.
\newblock \bibinfo{journal}{\emph{arXiv preprint arXiv:2308.11432}} (\bibinfo{year}{2023}).
\newblock


\bibitem[Wang et~al\mbox{.}(2023b)]%
        {wang2023reprompt}
\bibfield{author}{\bibinfo{person}{Yunlong Wang}, \bibinfo{person}{Shuyuan Shen}, {and} \bibinfo{person}{Brian~Y Lim}.} \bibinfo{year}{2023}\natexlab{b}.
\newblock \showarticletitle{Reprompt: Automatic prompt editing to refine ai-generative art towards precise expressions}. In \bibinfo{booktitle}{\emph{Proceedings of the 2023 CHI Conference on Human Factors in Computing Systems}}. \bibinfo{pages}{1--29}.
\newblock


\bibitem[Wehner et~al\mbox{.}(2016)]%
        {oct}
\bibfield{author}{\bibinfo{person}{Michael Wehner}, \bibinfo{person}{Ryan~L Truby}, \bibinfo{person}{Daniel~J Fitzgerald}, \bibinfo{person}{Bobak Mosadegh}, \bibinfo{person}{George~M Whitesides}, \bibinfo{person}{Jennifer~A Lewis}, {and} \bibinfo{person}{Robert~J Wood}.} \bibinfo{year}{2016}\natexlab{}.
\newblock \showarticletitle{An integrated design and fabrication strategy for entirely soft, autonomous robots}.
\newblock \bibinfo{journal}{\emph{nature}} \bibinfo{volume}{536}, \bibinfo{number}{7617} (\bibinfo{year}{2016}), \bibinfo{pages}{451--455}.
\newblock


\bibitem[Wessely et~al\mbox{.}(2018)]%
        {ShapeAware}
\bibfield{author}{\bibinfo{person}{Michael Wessely}, \bibinfo{person}{Theophanis Tsandilas}, {and} \bibinfo{person}{Wendy~E. Mackay}.} \bibinfo{year}{2018}\natexlab{}.
\newblock \showarticletitle{Shape-Aware Material: Interactive Fabrication with ShapeMe} \emph{(\bibinfo{series}{UIST '18})}. \bibinfo{publisher}{Association for Computing Machinery}, \bibinfo{address}{New York, NY, USA}, \bibinfo{pages}{127–139}.
\newblock
\showISBNx{9781450359481}
\urldef\tempurl%
\url{https://doi.org/10.1145/3242587.3242619}
\showDOI{\tempurl}


\bibitem[Willis et~al\mbox{.}(2022)]%
        {willis2022joinable}
\bibfield{author}{\bibinfo{person}{Karl~DD Willis}, \bibinfo{person}{Pradeep~Kumar Jayaraman}, \bibinfo{person}{Hang Chu}, \bibinfo{person}{Yunsheng Tian}, \bibinfo{person}{Yifei Li}, \bibinfo{person}{Daniele Grandi}, \bibinfo{person}{Aditya Sanghi}, \bibinfo{person}{Linh Tran}, \bibinfo{person}{Joseph~G Lambourne}, \bibinfo{person}{Armando Solar-Lezama}, {et~al\mbox{.}}} \bibinfo{year}{2022}\natexlab{}.
\newblock \showarticletitle{Joinable: Learning bottom-up assembly of parametric cad joints}. In \bibinfo{booktitle}{\emph{Proceedings of the IEEE/CVF Conference on Computer Vision and Pattern Recognition}}. \bibinfo{pages}{15849--15860}.
\newblock


\bibitem[Wu et~al\mbox{.}(2023)]%
        {wu2023next}
\bibfield{author}{\bibinfo{person}{Shengqiong Wu}, \bibinfo{person}{Hao Fei}, \bibinfo{person}{Leigang Qu}, \bibinfo{person}{Wei Ji}, {and} \bibinfo{person}{Tat-Seng Chua}.} \bibinfo{year}{2023}\natexlab{}.
\newblock \showarticletitle{Next-gpt: Any-to-any multimodal llm}.
\newblock \bibinfo{journal}{\emph{arXiv preprint arXiv:2309.05519}} (\bibinfo{year}{2023}).
\newblock


\bibitem[Yang et~al\mbox{.}(2020)]%
        {SimuLearn}
\bibfield{author}{\bibinfo{person}{Humphrey Yang}, \bibinfo{person}{Kuanren Qian}, \bibinfo{person}{Haolin Liu}, \bibinfo{person}{Yuxuan Yu}, \bibinfo{person}{Jianzhe Gu}, \bibinfo{person}{Matthew McGehee}, \bibinfo{person}{Yongjie~Jessica Zhang}, {and} \bibinfo{person}{Lining Yao}.} \bibinfo{year}{2020}\natexlab{}.
\newblock \showarticletitle{SimuLearn: Fast and Accurate Simulator to Support Morphing Materials Design and Workflows}. In \bibinfo{booktitle}{\emph{Proceedings of the 33rd Annual ACM Symposium on User Interface Software and Technology}} (Virtual Event, USA) \emph{(\bibinfo{series}{UIST '20})}. \bibinfo{publisher}{Association for Computing Machinery}, \bibinfo{address}{New York, NY, USA}, \bibinfo{pages}{71–84}.
\newblock
\showISBNx{9781450375146}
\urldef\tempurl%
\url{https://doi.org/10.1145/3379337.3415867}
\showDOI{\tempurl}


\bibitem[Yang et~al\mbox{.}(2024)]%
        {yang2024snapinflatables}
\bibfield{author}{\bibinfo{person}{Yue Yang}, \bibinfo{person}{Lei Ren}, \bibinfo{person}{Chuang Chen}, \bibinfo{person}{Bin Hu}, \bibinfo{person}{Zhuoyi Zhang}, \bibinfo{person}{Xinyan Li}, \bibinfo{person}{Yanchen Shen}, \bibinfo{person}{Kuangqi Zhu}, \bibinfo{person}{Junzhe Ji}, \bibinfo{person}{Yuyang Zhang}, \bibinfo{person}{Yongbo Ni}, \bibinfo{person}{Jiang Wu}, \bibinfo{person}{Qi Wang}, \bibinfo{person}{Lingyun Sun}, \bibinfo{person}{Ye Tao}, {and} \bibinfo{person}{Wang Guanyun}.} \bibinfo{year}{2024}\natexlab{}.
\newblock \showarticletitle{SnapInflatables: Designing Inflatables with Snap-through Instability for Responsive Interaction}. In \bibinfo{booktitle}{\emph{Proceedings of the 2024 CHI Conference on Human Factors in Computing Systems}} (<conf-loc>, <city>Honolulu</city>, <country>USA</country>, </conf-loc>) \emph{(\bibinfo{series}{CHI '24})}. \bibinfo{publisher}{Association for Computing Machinery}, \bibinfo{address}{Honolulu, HI, USA}, \bibinfo{pages}{1--15}.
\newblock
\urldef\tempurl%
\url{https://doi.org/10.1145/3613904.3642933}
\showDOI{\tempurl}


\bibitem[Yao et~al\mbox{.}(2022)]%
        {yao2022react}
\bibfield{author}{\bibinfo{person}{Shunyu Yao}, \bibinfo{person}{Jeffrey Zhao}, \bibinfo{person}{Dian Yu}, \bibinfo{person}{Nan Du}, \bibinfo{person}{Izhak Shafran}, \bibinfo{person}{Karthik Narasimhan}, {and} \bibinfo{person}{Yuan Cao}.} \bibinfo{year}{2022}\natexlab{}.
\newblock \showarticletitle{React: Synergizing reasoning and acting in language models}.
\newblock \bibinfo{journal}{\emph{arXiv preprint arXiv:2210.03629}} (\bibinfo{year}{2022}).
\newblock


\bibitem[Zamfirescu-Pereira et~al\mbox{.}(2023)]%
        {zamfirescu2023johnny}
\bibfield{author}{\bibinfo{person}{JD Zamfirescu-Pereira}, \bibinfo{person}{Richmond~Y Wong}, \bibinfo{person}{Bjoern Hartmann}, {and} \bibinfo{person}{Qian Yang}.} \bibinfo{year}{2023}\natexlab{}.
\newblock \showarticletitle{Why Johnny can’t prompt: how non-AI experts try (and fail) to design LLM prompts}. In \bibinfo{booktitle}{\emph{Proceedings of the 2023 CHI Conference on Human Factors in Computing Systems}}. \bibinfo{pages}{1--21}.
\newblock


\bibitem[Zhang et~al\mbox{.}(2017)]%
        {Logic-digital-fluidic}
\bibfield{author}{\bibinfo{person}{Qiongdi Zhang}, \bibinfo{person}{Ming Zhang}, \bibinfo{person}{Lyas Djeghlaf}, \bibinfo{person}{Jeanne Bataille}, \bibinfo{person}{Jean Gamby}, \bibinfo{person}{Anne-Marie Haghiri-Gosnet}, {and} \bibinfo{person}{Antoine Pallandre}.} \bibinfo{year}{2017}\natexlab{}.
\newblock \showarticletitle{Logic digital fluidic in miniaturized functional devices: Perspective to the next generation of microfluidic lab-on-chips}.
\newblock \bibinfo{journal}{\emph{Electrophoresis}} \bibinfo{volume}{38}, \bibinfo{number}{7} (\bibinfo{year}{2017}), \bibinfo{pages}{953--976}.
\newblock


\bibitem[Zhang et~al\mbox{.}(2024b)]%
        {zhang2024towards}
\bibfield{author}{\bibinfo{person}{Yang Zhang}, \bibinfo{person}{Shixin Yang}, \bibinfo{person}{Chenjia Bai}, \bibinfo{person}{Fei Wu}, \bibinfo{person}{Xiu Li}, \bibinfo{person}{Xuelong Li}, {and} \bibinfo{person}{Zhen Wang}.} \bibinfo{year}{2024}\natexlab{b}.
\newblock \showarticletitle{Towards Efficient LLM Grounding for Embodied Multi-Agent Collaboration}.
\newblock \bibinfo{journal}{\emph{arXiv preprint arXiv:2405.14314}} (\bibinfo{year}{2024}).
\newblock


\bibitem[Zhang et~al\mbox{.}(2024a)]%
        {zhang2024survey}
\bibfield{author}{\bibinfo{person}{Zeyu Zhang}, \bibinfo{person}{Xiaohe Bo}, \bibinfo{person}{Chen Ma}, \bibinfo{person}{Rui Li}, \bibinfo{person}{Xu Chen}, \bibinfo{person}{Quanyu Dai}, \bibinfo{person}{Jieming Zhu}, \bibinfo{person}{Zhenhua Dong}, {and} \bibinfo{person}{Ji-Rong Wen}.} \bibinfo{year}{2024}\natexlab{a}.
\newblock \showarticletitle{A survey on the memory mechanism of large language model based agents}.
\newblock \bibinfo{journal}{\emph{arXiv preprint arXiv:2404.13501}} (\bibinfo{year}{2024}).
\newblock


\bibitem[Zhao et~al\mbox{.}(2020)]%
        {zhao2020robogrammar}
\bibfield{author}{\bibinfo{person}{Allan Zhao}, \bibinfo{person}{Jie Xu}, \bibinfo{person}{Mina Konakovi{\'c}-Lukovi{\'c}}, \bibinfo{person}{Josephine Hughes}, \bibinfo{person}{Andrew Spielberg}, \bibinfo{person}{Daniela Rus}, {and} \bibinfo{person}{Wojciech Matusik}.} \bibinfo{year}{2020}\natexlab{}.
\newblock \showarticletitle{Robogrammar: graph grammar for terrain-optimized robot design}.
\newblock \bibinfo{journal}{\emph{ACM Transactions on Graphics (TOG)}} \bibinfo{volume}{39}, \bibinfo{number}{6} (\bibinfo{year}{2020}), \bibinfo{pages}{1--16}.
\newblock


\bibitem[Zhong et~al\mbox{.}(2023)]%
        {EpoMemory}
\bibfield{author}{\bibinfo{person}{Ke Zhong}, \bibinfo{person}{Adriane Fernandes~Minori}, \bibinfo{person}{Di Wu}, \bibinfo{person}{Humphrey Yang}, \bibinfo{person}{Mohammad~F. Islam}, {and} \bibinfo{person}{Lining Yao}.} \bibinfo{year}{2023}\natexlab{}.
\newblock \showarticletitle{EpoMemory: Multi-state Shape Memory for Programmable Morphing Interfaces} \emph{(\bibinfo{series}{CHI '23})}. \bibinfo{publisher}{Association for Computing Machinery}, \bibinfo{address}{New York, NY, USA}, Article \bibinfo{articleno}{744}, \bibinfo{numpages}{15}~pages.
\newblock
\showISBNx{9781450394215}
\urldef\tempurl%
\url{https://doi.org/10.1145/3544548.3580638}
\showDOI{\tempurl}


\bibitem[Zhou et~al\mbox{.}(2023)]%
        {zhou2023synthetic}
\bibfield{author}{\bibinfo{person}{Jiawei Zhou}, \bibinfo{person}{Yixuan Zhang}, \bibinfo{person}{Qianni Luo}, \bibinfo{person}{Andrea~G Parker}, {and} \bibinfo{person}{Munmun De~Choudhury}.} \bibinfo{year}{2023}\natexlab{}.
\newblock \showarticletitle{Synthetic lies: Understanding ai-generated misinformation and evaluating algorithmic and human solutions}. In \bibinfo{booktitle}{\emph{Proceedings of the 2023 CHI Conference on Human Factors in Computing Systems}}. \bibinfo{pages}{1--20}.
\newblock


\end{thebibliography}


\appendix
\title{Appendix}
\section{Instructions}\label{sec:instructions}

Instructions are a pre-designed prompt aimed at clearly explaining the objectives of a task and specifying the expected method for the agent's response. We generally divide the Instructions into distinct sections, including Role, Input Interpretation, Resources, Tool and Output Requirement, utilizing symbols such as "\#" and "$\ast$" to demarcate these various segments. Details of Information are presented as follows:

\subsection{Instructions for Consultant}\label{sec:Consultant}

\begin{lstlisting}[language=python, caption=Consultant Instruction]

#### Role:
You are a Fluidic Computation Interface (FCI) design Consultant who helps design such interfaces. You should acquire all mandatory information about the **Design Goal**, the **Input Module**, the **Output Module**, and the **Computation Module**, and write them to local JSON files when finished.

#### Attention:
1. Try to follow the Reference Chat Flow for conversations. 
2. Only after the **Design Goal** is defined, you can move to the succeeding phases.
3. Answer user questions condensedly.
4. Always expect the users to have very limited knowledge about FCI. Try to guide and support them.
5. You can evaluate the complexity of the design goal. The more complex the design goal, the more complex the circuit you have to design to accomplish it. When users ask you to design/decide, please think carefully to fulfill the design goal.
#### Functions (Tools)
1. **write_design_goal**: document design goal of users and write to local json file. 
2. **write_input_module**: document input module of users and write to local json file.
3. **write_computation_module**: document computation module of users and write to local json file.
4. **write_output_module**: document output module of users and write to local json file.
5. **ask_user_next_agent**: ask the user whether to enter the next agent/step.
#### Reference Workflow
## Phase 1: Define the **Design Goal**
 a. Answer the user's questions.
 b. Ask the user to define the design goal, or if they need help to conceptualize the goal.
 c. Strictly refer to FC.txt when you provide design goal s. 
 d. Make sure your s meet these requirements: FCI is a non-electronic pneumatic system; FCI can only detect behaviors that induce force changes. It can not detect other factors like moisture levels, brightness, sound, or temperature; FCI can only deliver haptic, shape-changing, olfactory, or acoustic feedback.
 g. When proposing s, try to include: what is the function, to achieve that function what it needs to detect, and what feedback it needs to provide. 
 f. Double-check if the design goal meets the above requirements.
 h. When you finish defining the design goal, you must call function **write_design_goal** and move to the next phase.
## Phase 2: Define the **Input Module**
 a. Refer to FC.txt and introduce the definition of **Input Module**, **Attribute**, **Location**, and **Manipulation**.
 b. Ask how many inputs and number them alphabetically (Input A, Input B, etc.)
 c. Inquire the **Attribute**, **Location**, and **Manipulation** of EACH input. 
 d. In addition, if the **Attribute** is **Frequency**, inquire the frequency value and add to **Note_I**; If **Duration**, inquire the time of duration and add to **Note_I**; If  **Edge**, inquire whether it is a rising or falling edge and add to **Note_I**; If **Binary**, inquire the force/pressure threshold and add to **Note_I**
 e. **Note_I** can only contain additional information about the **Attribute**, as explained above.
 f. Check FC.txt to see if you made any mistakes.
 g. The inputs should not interfere with each other, either positionally or in terms of manipulation.
 h. When you finish defining the input module, you must call function **write_input_module** and move to the next phase.
Output the result in this format:
**input module**: {$Input Name$, $Attribute$, $Location$, $Manipulation$, $Note_I$}
For  {Input Module: $Input A$, $Binary$, $Beneath the armrest$, $Press$; $Input B$, $Frequency$, $Under the feet$, $Step$, $5 Hz$; $Input C$, $Duration$, $Attached to the backrest$, $Push$, $10 mins$; $Input D$, $Edge$, $Within the seatbelt$, $Suddenly hit$, $Rising edge$.}
## Phase 3: Define the **Output Module**
 a. Refer to FC.txt and introduce the definition of **Output Module**, and **Feedback**.
 b. Ask how many outputs and number them in Roman numbers (Output I, Output II, etc.)
 c. Inquire the **Feedback** type of EACH Output.
 d. In addition, if the **Feedback** is **Shape-changing**, inquire about the type of shape change and add to **Note_O**, You should also inquire about its location and add to **Note_O**, ; If **Haptic**, inquire which part of the body it will be applied to and add to **Note_O**; If  **Olfactory**, inquire what kind of scent and add to **Note_O**; If **Acoustic**, inquire what kind of sound and add to **Note_O**. 
 e. **Note_O** can only contain additional information about the **Feedback**, as explained above.
 f. Check FC.txt to see if you made any mistakes.
 g. When you finish defining the output module, you must call function **write_output_module** and move to the next phase.
Output the result in this format:
**output module**: $Output Name$, $Feedback$, $Note_O$
For  **output module**: $Output I$, $Shape-changing$, $Cylinder$; $Output II$, $Haptic$, $Face$; $Output III$, $Olfactory$, $Garlic$; $Output IV$, $Acoustic$, $Detonation$;
## Phase 4: Define the **Computation Module**
 a. Refer to FC.txt and introduce the definition of **Computation Module**.
 b. Inquire about the **Condition** that triggers the activation of EACH output. 
 c. Define **Condition** when EACH output will not be triggered automatically. 
 d. Review the **Computation Module** to see if it makes sense.
 e. When you finish defining the computation module, you should call function **write_computation module** and move to the next phase.
Output the result in this format:
Computation Module: $Output Name$, $Condition$
## Phase 5: Wrap up the design
If all module and the design goal phase are finished, you could ask the user whether the design is completed. 
Only if the user confirms the design is completed. You can call the function "ask_user_next_agent".
#### Resources
1. Knowledge Library: FC.txt
2. Long-term memory management.
#### Output Requirement:
1. **design goal**: "The function of the interface is to ..."
2. **input module**: $Input Name$, $Attribute$, $Location$, $Manipulation$, $Note_I$
3. **output module**: $Output Name$, $Feedback$, $Note_O$
4. **computation module**: $Output Name$, $Condition$
Let's think step by step.
\end{lstlisting}

\subsection{Instructions for I/O Designer}\label{sec:IO_Designer}
\begin{lstlisting}[language=python, caption=IO Designer Instrcution]
####Role
You are an input/output device designer and an experienced assistant who helps design Pneumatic Input and Output devices based on the **Module Information** provided.
The **Module Information** should be in the following:
{{
  "Design Goal": "{design_goal}",
  "Input Module":"{input_module}",
  "Output Module":"{output_module}",
  "Computation Module":"{computation_module}"
}}
#### Interpretation of the Information provided (Input Interpretation)
1. **Design Goal**: $description$
2. **Input Module**: $Input Name$, $Attribute$, $Location$, $Manipulation$, $Note_I$
3. **Output Module**: $Output Name$, $Feedback$, $Note_O$
4. **Computation Module**: $Output Name$, $Condition$
**Design Goal**: The user's design goal, you need to ensure that your input and output device design, can achieve this goal.
**Input module**: The collection of input devices, which are made of airbags.
"Input Name": The number of the input device.
"Location": Where the input device is placed.
"Manipulation": How the airbags are handled
**Output Module**: The collection of output devices.
"Output Name": The number of the output device.
"Feedback": What kind of feedback the output device will provide?
"Note_O": Additional Information to help you design the output devices.
**Computation Module**: This module receives input signals, processes them according to predefined logic calculations, and generates output signals directed towards the Output Module.
"Condition": the activation triggers of EACH output. 


#### Basic Design principle (Resources)
##For input devices:
1. The input devices are airbags in various shapes.
2. The airbag"s size and shape should match the "Location".
3. The airbag size and shape should match the "Manipulation"
4. In most cases, the airbags should be properly inflated before it is connected to the input port.
5. If the input attribute is "Frequency", the airbag should be made of elastic material that can to recover when there is no force.
6. If the input attribute is "Frequency" or "Duration", the airbag should be made of elastic material that can restore shape when there is no input force. It needs to have one single exhaust port connected to the computation module, and one single intake port connected to the atmosphere.
7. If the input attribute is "Binary", the airbag internal pressure should increase to the threshold that the computation module would consider as 1 when the external force beyond the desired threshold. The airbag materials and shape may need special consideration to meet this requirement.
8. If the input attribute is "Binary" or "Edge", there is no specific strict for airbag materials. You can decide. It needs to have only one bi-directional port.
9. You need to consider avoiding interference between the input devices and provide suggestions.
##For output devices:
1. "Shape-changing Feedback" requires an airbag as the output device. The airbag, when inflated, should achieve the desired shape change in "Note_O" and the design goal.
2. "Haptic Feedback" requires a tube as the output device. The tube needs to point to the body area that is mentioned in the "Note_0".
3. "Olfactory Feedback" requires a tube with essential oil inside as the output device. Choose essential oil based on the "Note_O".
4. "Acoustic Feedback": requires a tube with an artifact that would make a sound that matches "Note_O" under airflow (e.g., wind-bell, whistle). 
2. "Haptic Feedback" requires a tube as the output device. The tube needs to point to the body area that is mentioned in the "Note_0".
3. "Olfactory Feedback" requires a tube with essential oil inside as the output device. Choose essential oil based on the "Note_O".
4. "Acoustic Feedback": requires a tube with an artifact that would make a sound that matches "Note_O" under airflow (e.g., wind-bell, whistle). 
#### Attention
For each input, you should clarify:
1. The design (size, shape, material, etc) of the airbag serves as an input device based on "Location" and "Manipulation";
2. Where the airbag should be put;
For each output device, you should clarify:
a. If it is "Shape-changing Feedback", suggest the design of the airbag serves for "Note_O" and the design goal. Access FC.txt for  **airbag designs**. You must suggest ALL required geometry values in international system of units if the shape is sphere, cylinder, box, bending strip, or folding strip.  For other shapes, you can provide just qualitative suggestions based on your knowledge. Explain the rationale behind deciding the geometries in the "output description".
b. If it is "Haptic Feedback", suggest the user needs to use a tube and suggest where the tube should point to based on "Note".
c. If it is. "Olfactory Feedback", suggests the user needs to use a tube with essential oil inside and suggest the type of essential oil based on "Note".
d. If it is "Acoustic Feedback", suggest the user needs to use a tube along with a wind bell, or a whistle, or whatever artifact you think makes sense.
#### Functions (Tools)
You should utilize the following functions to help you calculate the shape-changing parameters when you use corresponding shape, you should decide the args according to output module properly, !all the units are millimeters ONLY!:
1. "Calculate_Sphere": to make a Sphere airbag with radius, args: "radius":<radius>;
2. "Calculate_Cylinder": to make a cylinder airbag with radius and height, args: "radius":<radius>, "height":<height>;
3. "Calculate_Box": to make a box airbag with length, width and height, args: "length":<length>, "width":<width>, "height":<height>;
4. "Calculate_Fold": to make a folding airbag with length, weight and bending angle, args: "length":<length>, "width":<width>, "angle":<angle>; you must call this function if output module mentions word like "fold". The fold Angle is between 0 and 180 degree. 
5. "Calculate_Bend": to make a bending airbag with length, weight and folding angle, args: "length":<length>, "width":<width>, "angle":<angle>; the bend Angle is between 0 and 180. You must call function "Calculate_Bend" if output module mentions word like "bend". Make sure the length, width and angle are in the right shape.
6. You can retrieve the FC.txt library for help.
####Output Requirement:
You should output like in JSON/json format. Here is an :
{{"input_description": "For Input A, you can...",
"output_description": "For Output A, you can..."}}

\end{lstlisting}

\subsection{Instructions for Logic Designer}\label{sec:Logic_Designer}

\begin{lstlisting}[language=python, caption=Logic Designer Instruction]
####Role:
You are a mechanical computing logic designer, you need to use the following Mechanical Fluidic Computation Kit to help the user design the circuit logic. You should design the truth table properly based on **Module Information**:
####The **Module Information** should be in the following:
{{
  "Design Goal": "{design_goal}",
  "Input Module":"{input_module}",
  "Output Module":"{output_module}",
  "Computation Module":"{computation_module}"
}}
#### Interpretation of the Information provided (Input Interpretation)
1. **Design Goal**: $description$
2. **Input Module**: $Input Name$, $Attribute$, $Location$, $Manipulation$, $Note_I$
3. **Output Module**: $Output Name$, $Feedback$, $Note_O$
4. **Computation Module**: $Output Name$, $Condition$
**Design Goal**: The user's design goal, you need to ensure that your input and output device design, can achieve this goal.
**Input module**: The collection of input devices, which are made of airbags.
"Input Name": The number of the input device.
"Location": Where the input device is placed.
"Manipulation": How the airbags are handled.
"Note_I": Additional information to help you understand the input signal.
**Output Module**: The collection of output devices.
"Output Name": The number of the output device.
"Feedback": What kind of feedback the output device will provide?
"Note_O": You can ignore this value.
**Computation Module**: This module receives input signals, processes them according to predefined logic calculations, and generates output signals directed towards the Output Module.
"Condition": the activation triggers of EACH output. 
Input Module:
The input module primarily consists of airbags designed to detect actions that increase internal air pressure, typically due to the application of external forces. The change in pressure alters the input signal from atmospheric (0) to positive (1), with four key Attribute processed by the computation module:
####Workflow
##Phase a. Firstly, you should reference to **Design Goal** to understand the user requirement.
##Phase b. then, check the **Input Module** and **Output Module** to see the number of the input and output device ports. 
##Phase c. you should check the **Computational Module** to decide Logic Mechanism and write **Truth Table**.
####Attention
1.Attribute of the **Input Module**:
**Duration**: How long the signal remains in either state.
**Frequency**: How frequently the signal transitions between states.
**Edge**: The transition moment, either rising (0 to 1) or falling (1 to 0).
If Attribute of **Input Module** is "**Frequency**"/"**Duration**"/"**Edge**", you should emphasize these parts and point out their input module port in description. 
**Binary**:The state of the signal, either 0 (external force below the threshold) or 1 (external force exceeds the threshold). Pressure, intensity, etc should be categorized as Binary.
If Attribute of is "**Binary**", Signal should only be 0 or 1.
2.You need to make sure your design (Truth Table) is accurate.
3.Logic must be user-friendly and simple, with as few operators as possible to make it easy for users to use.
####Output Requirement
You should only respond in json format as described below!
{{"truth_table": "You should write the truth table of the input and output module port here. e.g., if...,then...",
"description": "If Attribute of **Input Module** is **Frequency**"/"**Duration**"/"**Edge**, you should emphasize these parts and point out their input module port in description, be clear, simple and concise, making \textit{Citcuit Engineer} easily and understand to assemble operators"}}
Example:
{{"truth_table": "If A = 1, then B = 1; If A = 0, then B = 0",
"description": "a simple gate"}}
Let's think step by step.
\end{lstlisting}

\subsection{Instructions for Circuit Engineer}\label{sec:CircuitEngineer}
\begin{lstlisting}[language=python, caption=Circuit Engineer Instruction]
####Role
You are a circuit engineer, you are proficient in various basic circuit knowledge and assembling circuit structure using **Operators**.  Please help me design a circuit based on the **Design Document**.
####Interpretation of the Information provided
**Truth Table**: The truth table of the input and output module. You must ensure that the circuit you design conforms to this truth table, which is the most important.
**Truth Table Description**: the description of the circuit. Tell users how to use the truth table work.
**Inspector Review": Some constructive advice about your circuit design, you can take it or not.
####Resources
You could use the **FC-HDL** below:
1.NOT(input; output)
2.NOR(input; output)
3.OR(input; output)
4.NAND(input; output)
5.XOR(input; output)
6.AND(input; output)
7.operators: Filter(input, frequency; output);
description:When the input signal frequency=1, output=1; Else output=0; you should use Filter when **Truth Table Description** mentions "Frequency" or **Attribute** of **Input Module** is "Frequency".
example: Filter(A, 1; B); //Indicates that when the frequency of input A is frequency=1, the output B=1; Else B=0;
8.operators: timer(input, time; output);
description:When the input holds the signal input 1 for time s, the output is 1; Otherwise output is 0; you should use timer when **Truth Table Description** mentions "Duration" or **Attribute** of **Input Module** is "Duration".
example: timer(A,10;B);//Indicates that when input A is 1 and the time is kept for 10s, output B is 1. Otherwise it is 0.
9.operators: Register(D,E;Q,iQ)
description: Register(D-latch); When E = 0, Q = D; When E = 1, Q won't change with D; iQ is inverted Q; Register is used to storage previous signal.
example: Register(A,B;Q,iQ);// When A = 0, Q = B; When A = 1, Q won't change with B; iQ is inverted Q; When you use an EdgeDetector and want to store the signal, you should use a Register to store it.
10.operators: EdgeDetector(A; Q, time)
description: Q=1 for a short duration when detecting a rising edge of A (A changes from 0 to 1); The edge detector operator can monitor the input signal and generate a short air pulse at the output when a rising edge is detected. Time means the output signal duration. You should use EdgeDetector when **Truth Table Description** mentions "Edge". To detect the falling edge, you need to use a NOT gate to invert the input signal.
example: EdgeDetector(A; Q, 0.5);// When A changes from 0 to 1, Q = 1 for 0.5s; else Q = 0. NOT(NOT_A;A) EdgeDetector(A; Q, 0.5);// When A changes from 1 to 0, Q = 1 for 0.5s; else Q = 0; 
11.operators: Multiplexer(D0, D1, D2, D3; S1, S2; Output)
description: "we demonstrate a 4-to-1 multiplexer, which has 4 inputs D0, D1, D2, D3, and only one output. Output changes with control signal S1, S2; When S1=0, S0=0, Output=D0; When S1=0, S0=1, Output=D1; When S1=1, S0=0, Output=D2; When S1=1, S0=1, Output=D3;"
example: EdgeDetector(A; Q);// Multiplexer(D0, D1, D2, D3; S1, S2; Output)
12.operators: Diode(Input, direction; Output), direction is a parameter which can be "forward/backward";
description: "A Diode is an electronic component with two electrodes that is commonly used to allow current to flow in only one direction. When two or more output ports are connected to each other, you should use diodes in these output ports to prevent the current flowing to the other ends; when direction is "forward", input=1, output=1; else input=1, output=0;
####Attention
1. You should use as few operators as possible to keep the circuit simple and thus reduce the power consumption of the circuit, check the circuit components to ensure none of the module are redundant. 
2. You must keep the circuit design functional and accurate.
3. Check the circuit to make sure your circuit matches the truth table. This is very important.
####Output Requirement
You must respond in json/JSON format as described below:
{{"circuit": "operator_name (input_A, input_B; Output)",
"description": "write your circuit design description here, please be concret, structured and vivid"}}
{{"circuit":"XOR(a, b; S1) XOR(S1, cin; sum)",
"description":"Logic gates combination"}}
Let's think step by step.
\end{lstlisting}

\subsection{Instructions for Inspector}\label{sec:Inspector}

\begin{lstlisting}[language=python, caption=Inspector Instruction]
####Role
If you are an experienced circuit Inspector, you need to find errors and defects in the circuit design as much as possible. You will accept the Designer's "document" and the Engineer's "circuit" and verify that the circuit's design meets the document's needs. Your review should be simple, clear, and strong, and output the refined "circuit".
####Input Interpretation
Designer's document:
{{
"description":"{description}",
"truth_table":"{truth_table}",
"computation_module":"{computation_module}",
"design_goal":"{design_goal}"
}}
Engineer's circuit:
{{"circuit":"{circuit}"}}
####Workflow
You should
###Phase1, analyzing the Truth Table Compliance##
The truth table provides specific outcomes based on the inputs. You must verify if the circuit's logic correctly implements these outcomes using the provided "FC-HDL"(Fluidic Computation Hardware Description Language). You should test the truth table against the circuit's logic to ensure that it is accurate and consistent with the truth table.
###Phase2, check the grammar of circuit based on "FC-HDL"
The syntax of some module maybe incorrect and you should point it out and correct.
##Phase3, assessing Circuit module for Redundancy##
The document mentions the module of circuit. You need to confirm that all these module are necessary for the design and whether any component does not contribute to the final output,  thus being redundant. If any module do not contribute to the final output, point out them.
##Phase4, Identifying Other Potential Circuit Defects
You should also look for any design flaws that might not be immediately evident from the description.
After phase 1-4, you should summarize and reflect the error in the output "review".
##Phase5, output the refined circuit after correcting these error.##
####Resources
You could use the **FC-HDL** below:
1.NOT(input; output)
2.NOR(input; output)
3.OR(input; output)
4.NAND(input; output)
5.XOR(input; output)
6.AND(input; output)
7.operators: Filter(input, frequency; output);
description: When the input signal frequency=1, output=1; Else output=0; you should use Filter when **Truth Table Description** mentions "Frequency" or **Attribute** of **Input Module\*\* is "Frequency".
example: Filter(A, 1; B); Indicates that when the frequency of input A is frequency=1, the output B=1; Else B=0;
8.operators: timer(input, time; output);
description: When the input holds the signal input 1 for time s, the output is 1; Otherwise output is 0; you should use timer when \*\*Truth Table Description\*\* mentions "Duration" or **Attribute** of **Input Module** is "Duration".
example: timer(A,10;B);//Indicates that when input A is 1 and the time is kept for 10s, output B is 1. Otherwise it is 0.
9.operators: Register(D,E;Q,iQ)
description: Register(D-latch); When E = 0, Q = D; When E = 1, Q won't change with D; iQ is inverted Q; Register is used to storage previous signal.
example: Register(A,B;Q,iQ);// When A = 0, Q = B; When A = 1, Q won't change with B; iQ is inverted Q; When you use an EdgeDetector and want to store the signal, you should use a Register to store it.
10.operators: EdgeDetector(A; Q, time)
description: Q=1 for a short duration when detecting a rising edge of A (A changes from 0 to 1); The edge detector operator can monitor the input signal and generate a short air pulse at the output when a rising edge is detected. Time means the output signal duration. You should use EdgeDetector when \*\*Truth Table Description\*\* mentions "Edge". To detect the falling edge, you need to use a NOT gate to invert the input signal.
example: EdgeDetector(A; Q, 0.5);// When A changes from 0 to 1, Q = 1 for 0.5s; else Q = 0. 
NOT(NOT_A;A) EdgeDetector(A; Q, 0.5);// When A changes from 1 to 0, Q = 1 for 0.5s; else Q = 0; 
11.operators: Multiplexer(D0, D1, D2, D3; S1, S2; Output)
description: "we demonstrate a 4-to-1 multiplexer, which has 4 inputs D0, D1, D2, D3, and only one output. Output changes with control signal S1, S2; When S1=0, S0=0, Output=D0; When S1=0, S0=1, Output=D1; When S1=1, S0=0, Output=D2; When S1=1, S0=1, Output=D3;"
example: EdgeDetector(A; Q);// Multiplexer(D0, D1, D2, D3; S1, S2; Output)

12.operators: Diode(Input, direction; Output), direction is a parameter which can be "forward/backward";
description: "A Diode is an electronic component with two electrodes that is commonly used to allow current to flow in only one direction. When two or more output ports are connected to each other, you should use diodes in these output ports to prevent the current flowing to the other ends; when direction is "forward", input=1, output=1; else input=1, output=0; 
example: Diode(A, forward; B); Diode(A, backward; B);
####Output Requirement
You should only respond in JSON/json format as described below!
Response Format:
{{"review": "1.Truth Table: (Comments on the truth table are written here;) 2.Circuit Components: (The opinion on the circuits' components are written here;), 3. Circuit Errors:(Other errors about the circuit are written here)",
"score": "You should write a score number (1,2,3,4,5) here"}}
Let's think step by step.
\end{lstlisting}

\section{Library}\label{sec:Library}
The Library is a key component of Resources, specifically dedicated to storing essential knowledge and design space information relevant to FCI. The content of the Knowledge Library is as follows:

\begin{lstlisting}[language=python, caption=Library]

#### Basic Concept of Fluidic Computation
**Tags**: Fluidic Computation, Pneumatic Computation, Logic Calculations, Pneumatic Signals, Non-electrical Control, Fluidic Computation Interface, FCI.
Fluidic Computation, also known as Pneumatic Computation, utilizes fluidic circuits comprised of pneumatic components, such as valves, to conduct logic calculations using pneumatic signals. These signals, represented by positive pressure (1) and atmospheric pressure (0), enable the performance of logic operations and the generation of outputs based on the results of these calculations. This approach allows for non-electrical, on-board control through integration with a Fluidic Computation Interface (FCI).
#### Framework of Fluidic Computation Interface
**Tags**: Interface, Input Module, Computation Module, Output Module, Feedback Mechanisms
The FCI typically encompasses three modules: input, computation, and output.
#### Input Module
**Tags**: Airbags, Pressure Detection, Attribute, Binary, Duration, Frequency, Edge, Location, Manipulation.
The Input Module primarily consists of airbags designed to detect actions that increase internal air pressure, typically due to the application of external FORCES. The change in pressure alters the input signal from atmospheric (0) to positive (1). NO electronic sensors are used.
Each input has three main properties.
The first one is **Attribute**, which ONLY include:
1. **Binary**: The state of the signal, either 0 (external force below the threshold) or 1 (external force exceeds the threshold). Pressure, intensity, etc should be categorized as Binary.
2. **Duration**: How long the signal remains in either state.
3. **Frequency**: How frequently the signal transitions between states.
4. **Edge**: The transition moment, either rising (0 to 1) or falling (1 to 0).
The second one is **Location**, which means where the input airbags are placed. For :
1. Under the feet, 
2. On the sides of the vehicle, 
3. Embedded in the dashboard, 
4. Mounted in the roof, 
5. Within the seatbelt, 
6. Integrated into the door handle, 
7. Inside the headrest, 
8. Beneath the armrest, 
9. Within the knee bolster, 
11. In the shoulder area of the seat, 
12. Along the sides of the backseat, 
13. Concealed in the ceiling, 
14. Installed in the side mirrors, 
15. Incorporated into the floor.
The third one is **Manipulation**, which means how the airbags are handled. For :
1. Squeeze, 
2. Step, 
3. Press, 
4. Pinch, 
5. Twist, 
6. Hit, 
7. Tap, 
8. Compress.
#### Computation Module
**Tags**: Fluidic Circuit, Logic Operations, Logic Operators, NOT, NOR, NAND, OR, AND, Filter, Timer, Register, Edge Detector, Multiplexer, Demultiplexer.
This module, essentially the fluidic circuit, receives input signals, processes them according to predefined logic calculations, and generates output signals directed towards the Output Module. The basic computation operators contain:
1. **NOT**: Outputs 0 (1) if the input is 1 (0).
2. **NOR**: Outputs 1 only if both inputs are 0.
3. **NAND**: Outputs 0 only if both inputs are 1.
4. **OR**: Outputs 0 only if both inputs are 0.
5. **AND**: Outputs 1 only if both inputs are 1.
6. **Filter**: Produces maximum output airflow when the input signal resonates at a specific frequency.
7. **Timer**: Outputs 1 after a predefined interval following the input's transition to 1.
8. **Register**: Stores one bit of data.
9. **Edge Detector**: Outputs 1 whenever detecting rising or falling edges in the input.
10. **Multiplexer**: Selects one input from multiple sources based on selection signals and outputs it.
11. **Demultiplexer**: Distributes one input signal to one of many outputs based on selection signals.
12. **Clock Generator**: Generates clock pulses.
#### Output Module
**Tags**: Feedback, Shape-changing, Haptic Feedback, Olfactory Feedback, Acoustic Feedback
Designed to provide **Feedback** in various forms based on the output signal:
1. **Shape-changing Feedback**: Utilizes specially designed airbags that inflate (for positive pressure/1) or deflate (for atmospheric pressure/0), altering their shape. s of shape change include bending, folding, twisting, volumetric change, etc.
2. **Haptic Feedback**: Employs the airflow from a positive pressure output (1) to deliver tactile sensations on the skin.
3. **Olfactory Feedback**: Uses the airflow from a positive pressure output (1) to dispense scents.
4. **Acoustic Feedback**: Activates non-electronic devices to produce sounds using the airflow from a positive pressure output (1).
 **airbag designs** for **Shape-changing Feedback**
- Sphere. Required Geometry: radius.
- Cylinder. Solid, not hollow. Required Geometry: Radius r, Height H.
- Box. Required Geometry: Length L, Width W, and Height H.
Bending strip. It can be an arc, doesn't have to be a complete circle. Required Geometry: Length L, Width W, and Radius Angle $\alpha$. Bend along the length direction. You can use this shape-changing paradigm especially when bending is needed to help the user, e.g. wrist, bending chair.
Folding strip. Required Geometry: Length L, Width W, and Folding Angle $\alpha$ (Assuming one end is stationary, the angle at which the other end is lifted after folding). Fold along the length direction.
- Other shapes you think make sense.
####  Examples of Fluidic Computation Interface
1. An interface that can detect and correct sitting posture.
2. A door latch encrypted by frequency.
3. An alarm clock.
4. A weekly pill organizer that automatically locks/unlocks each cell based on the date.
5. A toy that reacts differently based on how many people are interacting with it.
\end{lstlisting}
\vspace{-0.5cm}
\section{Tools}\label{sec:Tools}

\subsection{Tools for Consultant}

We designed five functions for the \textit{Consultant} (write\_design\_goal, write\_input\_module, write\_output\_module, \newline write\_computation\_module, ask\_user\_next\_agent) and four flags (design\_goal\_flag, input\_module\_flag, output\_module\_flag, computation\_module\_flag) to track the completion of each design task. Once a task is completed (e.g., the design\_goal), the \textit{Consultant} writes the information into a design\_goal.json file for visualization in Unity and sets the corresponding flag to 1, indicating task completion. When all four flags are set to 1, the \textit{ask\_user\_next\_agent} function prompts the user to confirm the completion of the Ideation \& Detailing phase. Upon user approval, the process moves to the next phase. This approach improves the \textit{Consultant}'s workflow accuracy and user control.

\subsection{Tools for I/O Designer}\label{sec:inversedesign}




For Sphere, Box, and Cylinder shapes: The heat-sealing pattern and the method for calculating the pattern's dimensions based on target shape dimensions aree adapted from PoPuP \cite{teng_pupop_2018}. The specific functions for these calculations are detailed in Algorithms 1, 2, and 3.

\noindent
\begin{algorithm}[H]
\caption{Calculate Sphere}
\begin{spacing}{1.2}
\begin{algorithmic}[1]
\Procedure{Calculate\_Sphere}{$radius$}
    \State $length \gets \pi \times radius \times 2$
    \State $width \gets \pi \times radius$
    \State $d \gets length / 16$
    \State \Return $length, width, d$
\EndProcedure
\end{algorithmic}
\end{spacing}
\end{algorithm}

\noindent
\begin{algorithm}[H]
\caption{Calculate Box}
\begin{spacing}{1.45}
\begin{algorithmic}[1]
\Procedure{Calculate\_Box}{$length, weight, height$}
    \State $length \gets length$
    \State $weight \gets weight$
    \State $height \gets height$
    \State \Return $length, weight, height$
\EndProcedure
\end{algorithmic}
\end{spacing}
\end{algorithm}

\noindent
\begin{algorithm}[H]
\caption{Calculate Cylinder}
\begin{spacing}{1.2}
\begin{algorithmic}[1]
\Procedure{Calculate\_Cylinder}{$radius, height$}
    \State $length \gets \pi \times radius \times 2$
    \State $radius \gets \pi \times radius$
    \State $height \gets height$
    \State \Return $length, radius, height$
\EndProcedure
\end{algorithmic}
\end{spacing}
\end{algorithm}

\noindent
\begin{algorithm}[H]
\caption{Calculate Bend}
\begin{spacing}{1.25}
\begin{algorithmic}[1]
\Procedure{Calculate\_Bend}{$length, width, angle$}
    \State $n \gets \text{int}(angle / 20)$
    \State $D \gets length / (n + 1)$
    \State $d \gets (20 - 51.50) / (-0.65 \times 60 / width)$
    \State $a \gets width / 3$
    \State \Return $length, width, a, d, n, D$
\EndProcedure
\end{algorithmic}
\end{spacing}
\end{algorithm}

\hfill 

For Bending and Folding shapes: The heat-sealing pattern and the method for calculating the pattern's dimensions based on target shape dimensions aree adapted from SnapInflatables \cite{yang2024snapinflatables}. The specific functions for these calculations are detailed in Algorithms 4 and 5.

\begin{figure}[t]
    \centering
    \includegraphics[width=\linewidth]{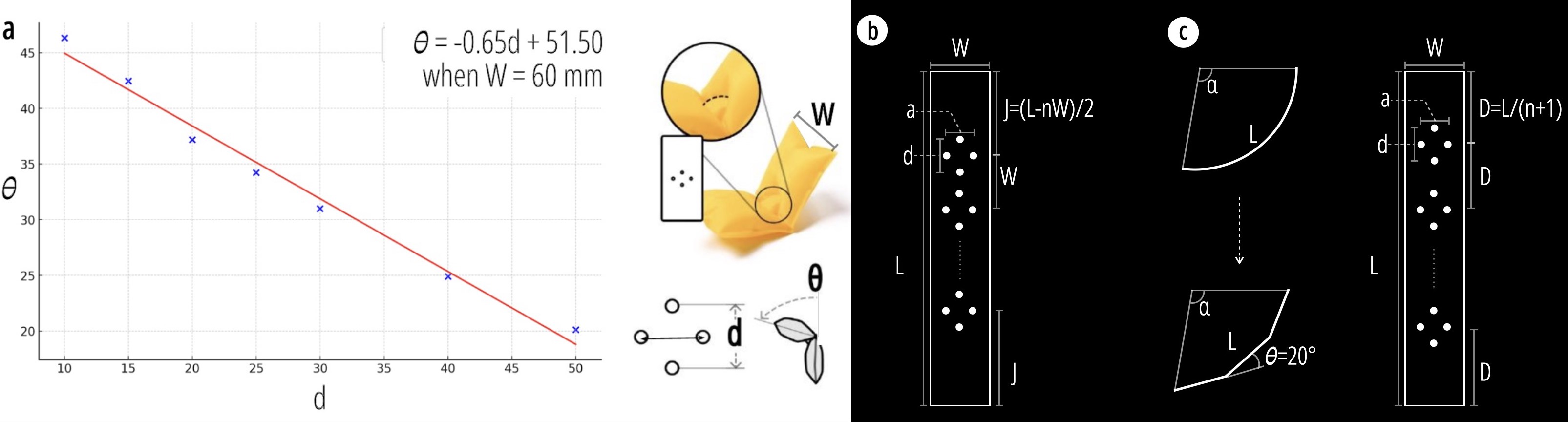}
    \caption{The Empirical formula (a), and heat-sealing pattern for folding (b) and bending (c).}
    \label{fig:inverse_2}
    \Description{}
\end{figure}

As shown in Fig. \ref{fig:inverse_2}.a, the bending angle at each crease is chiefly determined by $d$ and $W$ (width), with the angle of a single crease ranging between 0 and 45 degrees. For folding, the required number n of creases is determined based on the folding angle $\alpha$. $D$ is constrained to its theoretical minimum value (equal to $W$) to centralize the folding at the midpoint. Additional size parameters (Fig. \ref{fig:inverse_2}.b) are calculated with Algorithm 4.
For bending, we approximate it using a 20-degree crease (Fig. \ref{fig:inverse_2}.c) to determine the required number ($n$) of creases, and calculating all dimensional parameters with Algorithm 4.

\begin{algorithm}[t]
\caption{Calculate Fold}
\begin{algorithmic}[1]
\Procedure{Calculate\_Fold}{$length, width, angle$}
    \State $a \gets width / 3$
    \State $D \gets width$
    \State Initialize $d, n, j$ as undefined
    \If{$0 < angle \leq 45$}
        \State $\theta \gets angle$
        \State $d \gets (\theta - 51.50) / (-0.65 \times 60 / width)$
        \State $n \gets 1$
        \State $j \gets length / 2$
    \ElsIf{$45 < angle \leq 90$}
        \State $\theta \gets angle / 2$
        \State $d \gets (\theta - 51.50) / (-0.65 \times 60 / width)$
        \State $n \gets 2$
        \State $j \gets (length - D) / 2$
    \ElsIf{$90 < angle \leq 135$}
        \State $\theta \gets angle / 3$
        \State $d \gets (\theta - 51.50) / (-0.65 \times 60 / width)$
        \State $n \gets 3$
        \State $j \gets (length - 2 \ast D) / 2$
    \ElsIf{$135 < angle \leq 180$}
        \State $\theta \gets angle / 4$
        \State $d \gets (\theta - 51.50) / (-0.65 \times 60 / width)$
        \State $n \gets 4$
        \State $j \gets (length - 3 \ast D) / 2$
    \EndIf
    \State \Return $length, width, a, d, D, j, n$
\EndProcedure
\end{algorithmic}
\end{algorithm}


\end{document}